\documentclass[12pt,reqno]{article}

\usepackage[hmargin=2cm,bmargin=2cm,tmargin=2.5cm]{geometry}

\usepackage[T1]{fontenc}
\usepackage{enumitem,kantlipsum}
\usepackage[leqno]{amsmath}
\usepackage{amsthm}
\usepackage{amsxtra}
\usepackage{amsfonts}
\usepackage{amssymb}
\usepackage{color,hyperref}
\usepackage{graphicx}
\usepackage{tikz}
\usepackage{subfig}
\usepackage{cite}
\hypersetup{colorlinks,breaklinks,
linkcolor=blue,urlcolor=blue,
anchorcolor=blue,citecolor=blue}
\usepackage{color}
\usepackage{array}
\usepackage{booktabs}
\usepackage{dsfont}
\usepackage[toc,page]{appendix}
\usepackage{authblk}
\usepackage{booktabs}
\usepackage[font=small]{caption}
\usepackage{microtype}
\usepackage{nicefrac}
\usepackage[T1]{fontenc}
\usepackage{lmodern}
\usepackage{authblk}

\usepackage{graphicx}
\usepackage{color}
\usepackage{enumerate}
\usepackage{enumitem,kantlipsum}
\usepackage[english]{babel}

\usepackage{amsmath,amssymb}
\usepackage{bm}
\usepackage{amsthm}
\usepackage{mathtools}
\usepackage[utf8]{inputenc}
\usepackage[T1]{fontenc}
\usepackage[b]{esvect}
\usepackage{amssymb}
\usepackage{algorithmicx}
\usepackage[ruled]{algorithm}
\usepackage{algpseudocode}
\usepackage{parskip}
\usepackage{enumerate}
\usepackage{bm}
\usepackage{array}
\usepackage{arydshln}
\usepackage{multirow}
\usepackage{bigstrut}
\usepackage{bbm}
\usepackage{esdiff}
\usepackage{cleveref}
\usepackage{caption}

\usepackage{etoolbox}
\patchcmd{\thebibliography}{\section*{\refname}}{}{}{}

\usepackage{relsize}
\usepackage{environ}
\NewEnviron{myequation}{%
	\begin{equation*}
	\scalebox{.825}{$\BODY$}
	\end{equation*}
}
\NewEnviron{myequation1}{%
	\begin{equation*}
	\scalebox{.925}{$\BODY$}
	\end{equation*}
}
\NewEnviron{myequation2}{%
	\begin{equation*}
	\scalebox{.925}{$\BODY$}
	\end{equation*}
}
\NewEnviron{myequation3}{%
	\begin{equation*}
	\scalebox{.875}{$\BODY$}
	\end{equation*}
}
\newtheorem{thm}{Theorem}[section]
\newtheorem{lem}[thm]{Lemma}
\newtheorem{prop}[thm]{Proposition}
\newtheorem{cor}[thm]{Corollary}

\newtheorem{assum}{Assumption}

\newtheorem*{prob}{Problem Statement}

\theoremstyle{definition}
\newtheorem{defn}{Definition}
\newtheorem{exmp}{Example}
\newtheorem{remark}[thm]{Remark}

\newcommand{\cc}[1]{\overline{#1}}

\newcommand{\cA}{\mathcal{A}}
\newcommand{\cB}{\mathcal{B}}

\newcommand{\cD}{\mathcal{D}}
\newcommand{\cE}{\mathcal{E}}

\newcommand{\cG}{\mathcal{G}}
\newcommand{\cH}{\mathcal{H}}

\newcommand{\cK}{\mathcal{K}}
\newcommand{\cL}{\mathcal{L}}

\newcommand{\cP}{\mathcal{P}}
\newcommand{\cQ}{\mathcal{Q}}

\newcommand{\cS}{\mathcal{S}}

\newcommand{\cU}{\mathcal{U}}
\newcommand{\cV}{\mathcal{V}}

\newcommand{\bC}{\mathbb{C}}

\newcommand{\bR}{\mathbb{R}}

\newcommand{\vt}{\mathfrak{v}}

\DeclareMathOperator{\Dom}{Dom}

\def\up#1{^{(#1)}}

\title{On Vertex Conditions In Elastic Beam Frames: \\ Analysis on Compact Graphs}

\author{Soohee Bae}
\affil{Departmenet of Mechanical Engineering \\ Northeastern University}
\author{Mahmood Ettehad\thanks{corresponding author}}
\affil{Institute For Mathematics and Its Applications (IMA) \\ University of Minnesota}

\date{}

\begin{document}
	\maketitle
	
	\begin{abstract}
	We consider three-dimensional elastic frames constructed out of Euler-Bernoulli beams and describe extension of matching conditions by relaxing the vertex-rigidity assumption and the case in which concentrated mass may exists. This generalization is based on coupling an (elastic) energy functional in terms of field's discontinuities at a vertex along with purely geometric terms derived out of first principles. The corresponding differential operator is shown to be self-adjoint. Although for planar frames with a class of rigid-joints the operator decomposes into a direct sum of two operators, this property only holds for a special class of the proposed model. Application of theoretical results is then discussed in details for compact frames embedded in Euclidean spaces with different dimensions. This includes extension of the established results for rigid-joint case on exploiting the symmetry present in a frame and decomposing the operator by restricting it onto reducing subspaces corresponding to irreducible representations of the symmetry group. Derivation of characteristic equation based on the idea of geometric-free local spectral basis and enforcing geometry of the graph into play by an appropriate choice of the coefficient set will be discussed. Finally, we prove the limit conditions in parameter space which results in decomposing of vector-valued beam Hamiltonian to a direct sum of scalar-valued ones.    	
\end{abstract}

\section{Introduction}
\label{sec:Intro}
\begin{figure}[ht]
	\centering
	\includegraphics[width=0.7\textheight]{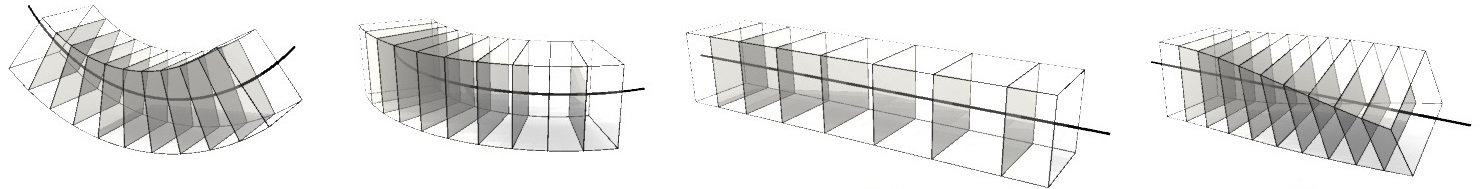}
	\caption{Degrees of freedom associated to an Euler-Bernoulli beam. This includes respectively (two) lateral displacements $v(x),  w(x)$, axial displacement $u(x)$, and angular (torsion) displacement $\eta(x)$.}
	\label{fig:dofs}
\end{figure}

Lattice materials are cellular structures obtained by tessellating a unit cell comprising a few beams or bars. Recently, the meta-material concept has been extended to materials showing novel mechanical behavior, e.g., the so-called auxetic materials exhibit the unusual mechanical property of having a negative Poisson’s ratio opening novel applications in molecular scales of crystallizing systems and to much larger scales in sound damping structures by band-gap opening and so on \cite{YLSXY04, CL13, KL15, ZEKBSA17, RDTNX18, LTWZ19}. From theoretical point of view, elastic deformation of continuous bodies maybe generally studied using the theory of elasticity. However, simplification on analysis can be reached by utilizing the kinematics of deformation and making some assumptions on the resulting strains. Under such simplification assumptions, modeling variety of natural and engineered tessellated lattices can generally be studied under beam theories \footnote{Most inclusive classical beam models are the Euler-Bernoulli and Timoshenko beam theories.}. Under Euler-Bernoulli beam model, e.g. see \cite{GR15}, each beam is described by an energy functional which involves four degrees of freedom for every infinitesimal element along the beam: lateral (2 degrees of freedom), axial, and angular displacements, see Figure \ref{fig:dofs}. At a joint, these four functions, supported on the beams involved, must be related via (matching) transmission conditions that take into account the physics of a joint, see \cite{BL04, GLL17, BE21} for more details. All of these make modeling of such structures made of joined together beams a topic of natural interest for engineering research and, more recently, for mathematicians working on differential equations defined on special metric spaces, namely quantum graphs \cite{MehBelNic_eds01, KKU15, GM20, BE21}.  This is equivalent to modeling the lattice as a discrete graph with set of vertices and edges joining them in physical space. More formally, a quantum graph is a metric graph equipped with a differential operator “Hamiltonian” and suitable vertex matching conditions. We refer to the work \cite{BK13} for detail background along this line.  

Recently, extension of the results to higher-order Hamiltonian gains interest by the more theoretically oriented communities. As an example, in the work \cite{KKU15, ZP16} vertex conditions for differential operator of fourth derivative on the planar metric graph are presented. It is shown that such conditions corresponding to the free movement of beams and depending on the angles among them in the equilibrium state. Non-trivial extension of this result to the three dimensional case is presented in \cite{BE21} in which joint conditions has been derived out of first principles. This extension is mainly based on coupling all degrees of freedom at the so called rigid-joint, i.e., joint in which the displacement and rotation vector fields are remained continuous at vertex set. It is shown that for rigid vertex model, field's coupling is inevitable even in the planar lattice case. This makes the corresponding Hamiltonian a vector valued self-adjoint operator on graph with physically interpretable vertex conditions, namely, equilibrium of net forces and moments applied on the vertex.  
\begin{prob}
	\label{probLems}
	Main purpose of current manuscript is to answer the following problems\footnote{These questions have been raised in the outlook Section of recent work \cite{BE21}.}
	\begin{itemize}
		\label{probm1}
		\item [(i)] how to generalize vertex model by relaxing the rigidity assumption so that discontinuity of fields are admissible in the solution space? 
		\label{probm2}
		\item [(ii)] under which (parametric) limits the vector-valued beam Hamiltonian on planar graphs decomposes to a direct sum of scalar-valued ones? 
	\end{itemize}
\end{prob}
Answering the problems above will provide a more complete picture on analysis of such continua. Most activities along this line have been focused on applied directions, while more theoretical works are mainly limited to {S}chr{\"o}dinger-type operator or higher order operators applied on simple geometry, e.g. a serially-connected beams connected by vibrating point masses \cite{B85, LLS93, CZ00, DN00, LR03, MR08, Kuc_bams16,ZP16,KM21}. Thereby, this manuscript is written with a hope of accommodating a step further towards physically sound analysis of observed phenomenon in the recently developed (meta) materials, and understanding the role of vertex transmission conditions on spectrum of Hamiltonian defined on metric graphs, see Section \ref{sec:Outlook} for potential extension and application of the current work. 

This paper is structured as follows: in Section \ref{sec:Elasticity}, after some preliminary discussion, we put forward a simple, general and compact description of the generalized semi-rigid vertex model. Definition of this type of joint is based on triggering both geometric description of the deformed frame (similar to the rigid-joint model) as well as inclusion of an energy functional in terms of displacement and rotation discontinuities, see Definition \ref{def:semiRigid}. In a special case of vanishing energy at the vertex, our result will be equivalent to the rigid-joint model derived out of purely geometric first principles in \cite{BE21}, see Definition \ref{def:rigid}. Next, formal derivation of self-adjoint Hamiltonian on graph with corresponding vertex conditions will be presented, see Theorem \ref{MainThm}. This will be followed by decomposition of the Hamiltonian on planar graphs, see Corollary \ref{decouplingPlane}. In Section \ref{sec:CompactGraph}, we discuss in details application of our theoretical framework on two compact frames, namely cantilevered beam and a three dimensional example in which the symmetry can be exploited to decompose the operator by restricting it onto reducing subspaces corresponding to irreducible representations of the symmetry group, see Theorem \ref{thm:reducing_antenna_tower}. Focus of Section \ref{sec:CompactGraph2D} will be on spectral analysis of 3-star planar finite graphs. This will be done by introducing set of local spectral basis defined on an interval $[0,1]$ followed by derivation of purely geometric basis. These two unrelated quantities then will be combined to construct characteristic equation corresponding the eigenvalue problem on such class of graphs, see Proposition \ref{Prop2D_1}. This concludes answering question (i) in problem statement, see Theorem \ref{MainThm}. At the end of last Section we will formally answer question (ii) in the problem statement, see Theorem \ref{thm:decouplingThm}. We conclude this manuscript by presenting a partial list of
directions for further mathematical investigations together with additional references.  
\section{Elasticity on Beam Frame}	
	\label{sec:Elasticity}
	\subsection{Preliminary}
	\label{sec:Preliminary}
	In this Section we first briefly review parameterization of a beam's elastic deformation. This requires introducing two vector valued quantities, namely displacement and rotation vectors defined out of geometric description of beams, followed by recalling the notation of a rigid-vertex, see definition \ref{def:rigid}, at which edges are met and has been developed in former work \cite{BE21}. Main contribution in this Section will be on generalization of the rigid-vertex model to a semi-rigid one, see definition \ref{def:semiRigid}. 
	\subsubsection{\textbf{Parameterization of Beam Deformation}} According to the Euler--Bernoulli hypothesis, which states that “plane sections remain plane,” the geometry of the spatial beam is described by the centroid line\footnote{The term \textit{centroid} indicates that this line is the locus of the centers of mass of the cross-sections.} and a family of the corresponding cross-sections. A fixed spatial basis with orthonormal base vectors $\{\vec E_1, \vec E_2, \vec E_3\}$ is introduced, which span the physical space (three dimensional Euclidean space) in which the beam is embedded. Moreover a family of orthonormal basis $\{\vec i,\vec j, \vec k\}$, called the cross-section basis or the material basis is employed to describe the orientation of the cross section of beam. The deformed configuration of the beam can be fully described by the position vector $\vec g(x)$ with $x$ representing the arc-length coordinate of the reference configuration, along with the family of orthonormal basis $\{\bm{\vec i}(x),\bm{\vec j}(x), \bm{\vec k}(x)\}$ which describe the orientation of the cross sections in the deformed configuration, see \cite{BE21} for detail setup. The displacement vector $\vec{g}(x)$ in the reference basis of the undeformed beam decomposes as 
	\begin{equation}
		\label{eq:dispComp}
		\vec{g}(x) := u(x)\vec{i} + w(x)\vec{j} + v(x)\vec{k}.
	\end{equation}
	The component $u(x)$ is called the \textit{axial displacement} while
	$w(x)$ and $v(x)$ are \textit{lateral displacements}. Introducing \textit{in-axis angular displacement} of the form $\eta(x) := \bm{\vec{j}}(x) \cdot \vec{k}$, then for small deformation regime, linear part of rotation vector  $\vec \omega(x)$ has representation, see Lemma 3.3 in \cite{BE21} for this derivation 
	\begin{equation}
	\label{eq:omega_lin}
	\vec \omega(x) = \eta(x) \vec i - v'(x) \vec j + w'(x) \vec k
	\end{equation}
	Consider now several beams, labeled $e_1, \ldots, e_n$ coupled at a joint. The type of matching vertex conditions there has a central role on type of coupling among the edges and thereby on global characteristic of the frame. Next, we will generalize the notion of rigid-joint model in \cite{BE21}.  
	\subsubsection{\textbf{Generalization of Joint Model}} 	A beam frame can be described as a geometric graph $\Gamma = (\cV,\cE)$, where $\cV$ denotes the set of vertices and $\cE$ the set of edges. The vertices $\vt \in \cV$ correspond to joints and edges $e \in \cE$ are the beams. Each edge $e$ is a collection of the following information: origin and terminus vertices $\vt_{e}^{o},\vt_{e}^{t}\in \cV$, length $\ell_e$ and the local basis $\{\vec{i}_e, \vec{j}_e, \vec{k}_e\}$. Describing the vertices $\cV$ as points in $\bR^3$ also fixes the length $\ell_e$ and the axial direction $\vec{i}_e$ (from origin to terminus). We will use the \textit{sign indicator} $s_e^{\vt}$ which is defined to be $-1$	when $\vt = \vt_{e}^{o}$, and $+1$ if $\vt = \vt_{e}^{t}$ and $0$ otherwise. This sign convention is consistent with the sign of outward normal derivative vector of an external surface applied in continuum mechanics literature, e.g. see \cite{ER15}. In this paper an edge $e$ incident to $\vt$ will be denoted by notation $e \sim \vt$, and moreover set  
	\begin{equation}
		\label{eq:vertexLim}
		\vec g_e(\vt) := \lim_{x\to\vt} \vec g_{e}(x), \qquad
		\vec \omega_e(\vt) :=  \lim_{x\to\vt} \vec \omega_{e}(x).
	\end{equation}  
	Associated to each vertex $\vt$ are two unknown vectors $\vec g_\vt^{\hspace{0.5mm} \circ}  \in \bC^3$ and $\omega_\vt^{\hspace{0.5mm} \circ} \in \bC^3$ with their components written with respect to the global coordinate system $\{\vec E_1, \vec E_2, \vec E_3\}$. Next we will define two class of vertex matching conditions which posses a central role in the remaining sections.   
	\begin{defn}
		\label{def:rigid}
		A joint $\vt$  is called
		\textit{rigid}, if for all $e \sim \vt$ the displacement and rotation on beams 
		satisfy continuity conditions 
		\begin{subequations}
			\label{rigidVertexDefn}
			\begin{gather}
				\label{eq:rigid_displacement}
				u_e(\vt) \vec i_e + w_e(\vt) \vec j_e +v_e(\vt) \vec k_e = \vec g_\vt^{\hspace{0.5mm} \circ}  \\
				\label{eq:rigid_rotation}
				\eta_e(\vt) \vec i_e - v_e'(\vt) \vec j_e + w_e'(\vt) \vec k_e  =  \vec \omega_\vt^{\hspace{0.5mm} \circ}
			\end{gather}
		\end{subequations}
	\end{defn}
	Observe that rigid-vertex model in the Definition \ref{def:rigid} is derived based on purely geometric description of a frame. However, in order to extend this definition to a semi-rigid one, it will be required to include energies developed due to discontinuity of fields at the vertex set. Let, denote by $S^n_+$ to be set of real non-negative valued symmetric positive definite matrices of size $n$.
	\begin{defn}
		\label{def:semiRigid}
		A joint $\vt$ with assigned set of (stiffness) matrices $\{\cK_{g_e}\up{\vt}\}_{e\sim \vt}$ and $\{\cK_{\omega_e}\up{\vt}\}_{e\sim \vt}$ elements of $S_+^3$ is called
		\textit{semi-rigid}, if for each edge $e \sim \vt$ it satisfies
		\begin{subequations}
			\label{eq:semirigidVertexDefn}
			\begin{gather}
				\label{eq:semiRigid_displacement}
				u_e(\vt) \vec i_e + w_e(\vt) \vec j_e +v_e(\vt) \vec k_e + s_e^\vt (\cK_{g_e}\up{\vt})^{-1} (c_e u_e'(\vt) \vec i_e - b_e w_e'''(\vt) \vec j_e
				- a_e v_e'''(\vt) \vec k_e) = \vec g_\vt^{\hspace{0.5mm} \circ}  \\
				\label{eq:semiRigid_rotation}
				\hspace{8mm}\eta_e(\vt) \vec i_e - v_e'(\vt) \vec j_e + w_e'(\vt) \vec k_e  + s_e^\vt (\cK_{\omega_e}\up{\vt})^{-1}(d_e \eta_e'(\vt) \vec i_e - a_e v_e''(\vt) \vec j_e +  b_e w_e''(\vt) \vec k_e) \hspace{1.5mm} = \vec \omega_\vt^{\hspace{0.5mm} \circ}
			\end{gather}
		\end{subequations}
	\end{defn}
   	We stress that presence of higher-order derivative in the definition of semi-rigid joint model in \eqref{eq:semirigidVertexDefn} compare to the rigid one in \eqref{rigidVertexDefn} has a direct connection to the energy at the vertex due to discontinuity of displacement and rotation vectors. In fact, let denote by set of matrices $\{\cB_e\}_{e \in \cE}$ with $\cB_e \in \text{SO}(3)$ transforming vector presentation in global coordinate system to the local one associated to an edge $e$. Moreover, let denote by 
   	\begin{equation}
   	\label{eq:FeMe}
   	\begin{split}
   	\vec f_e(\vt) := s_e^\vt (c_e u_e' \vec i_e - b_e w_e''' \vec j_e - a_e v_e''' \vec k_e), \qquad 
   	\vec m_e(\vt) := s_e^\vt (d_e \eta_e' \vec i_e - a_e v_e'' \vec j_e + b_e w_e'' \vec k_e)
   	\end{split}
   	\end{equation}
   	to be force and movement vectors corresponding to an edge $e \sim \vt$ evaluated (in the limit sense) at $\vt$, then the corresponding developed force (moment) has a (linear-spring) form 
    \begin{equation*}
    		\vec f_e(\vt) = \cK_{g_e}\up{\vt} (\cB_e \vec g_\vt^{\hspace{0.5mm} \circ}-\vec g_{e}(\vt)), 
    		\qquad 
    		\vec m_e(\vt) = \cK_{\omega_e}\up{\vt}(\cB_e \vec \omega_\vt^{\hspace{0.5mm} \circ}-\vec \omega_{e}(\vt))
    \end{equation*} 
    Sum of these (or net) forces and moments, classically appear from operator-theoretic derivations to guarantee self-adjointness of the resulting operator which contains physically well-known interpretation, i.e. balance of net force and moments. Interested reader may refer to the work \cite{BE21} for detail discussion on derivation of self-adjoint operator corresponding rigid-joint model. However, in the semi-rigid case, individual force and moments corresponding to each edge appears directly in the definition of vertex model. Physical interpretation of each constituent terms in \eqref{eq:semirigidVertexDefn} are as follows: 
    \begin{itemize}
    \item[(i)] In condition \eqref{eq:semiRigid_displacement}, $c_e u_e'$ is the force developed in direction $\vec i_e$ due to in-axis tension of the edge $e$, while $a_e v_e'''$ and $b_e w_e'''$ are shear forces developed inside the edge in the directions $\vec k_e$ and $\vec j_e$, respectively.
    \item[(ii)] In condition \eqref{eq:semiRigid_rotation}, $d_e \eta_e'$ is the moment associated with angular displacement developed in direction $\vec i_e$ due to in-axis rotation of the edge $e$, while $a_e v_e''$ and $b_e w_e''$ represent bending moments of the edge $e$ in the directions $\vec j_e$ and $\vec k_e$, respectively.    
    \end{itemize}
    We finally remark that the vector quantities in  \eqref{eq:semiRigid_displacement}--\eqref{eq:semiRigid_rotation} remain invariant under local change in edge's coordinate system. Next, we will present energy form on graph and its corresponding Hamiltonian for a generally three dimensional frames with semi-rigid joint model specified above. 
	\subsection{Variational and Differential Formulations}
	\label{sec:Section3}
    In the context of the kinematic Euler--Bernoulli assumptions for beam, no pre-stress, or external force, the strain energy of the
    an edge $e$ is expressed as
    \begin{equation}
    \label{eq:energyFunctional}
    \cU^{(e)}(x) = \frac{1}{2} \int_e \big(
    a_e(x) |v''_e(x)|^2 +
    b_e(x) |w''_e(x)|^2 + c_e(x)|u'_e(x)|^2 +
    d_e(x)|\eta'_e(x)|^2 \big) dx
    \end{equation} 
    The integration here is over the beam $e$, parameterized by the arc-length $x\in[0,\ell_e]$. Throughout the rest of manuscript we assume each beam in our frame is homogeneous in the axial direction and hence $a_e(x) \equiv a_e$ and so on. Extension of all results to variable stiffness is straightforward. Regarding energy at the vertex set, let $\{\cK_{g_e}\up{\vt}\}_{e\sim \vt}$ and $\{\cK_{\omega_e}\up{\vt}\}_{e\sim \vt}$ be a family of stiffness-matrices in $S_+^3$ associate to edges $e\sim \vt$, then energy at $\vt$ due to the discontinuity of displacement and rotation fields is a functional of a form   
    \begin{equation}
    \label{eq:vertexEnergyFunctional}
    \cU^{(\vt)} = 
    \frac{1}{2} \sum_{e \sim \vt} \big(\|\cB_e \vec g_\vt^{\hspace{0.5mm} \circ}-\vec g_{e}(\vt) \|_{\cK_{g_e}\up{\vt}}^2+
    \|\cB_e \vec \omega_\vt^{\hspace{0.5mm} \circ}-\vec \omega_{e}(\vt)\|_{\cK_{\omega_e} \up{\vt}}^2\big)
    \end{equation}
    Above, we use the convention of weighted induced norm in which for a vector $\vec u = u_k \vec\iota_k$ and a matrix $\cK = \kappa_{ij} \vec \iota_i (\vec \iota_j)^T$ defined with respect to an orthonormal basis $(\vec \iota_1, \vec \iota_2, \vec \iota_3)$, then $\|\vec u\|_\cK^2 = \langle \vec u,\vec u \big\rangle_{\cK} =  \bar u_i \kappa_{ij} u_j$. Combining energy functional corresponding to the deformation on edge set $\cE$ and discontinuity at vertex set $\cV$ respectively defined in \eqref{eq:energyFunctional} and \eqref{eq:vertexEnergyFunctional}, then total energy of the beam is expressed as
	\begin{equation}
	\label{eq:energyFunctionalGraph}
		\cU\up{\Gamma} := \sum_{e \in \cE} \cU^{(e)} + \sum_{\vt \in \cV} \cU^{(\vt)}
	\end{equation}
	\subsubsection{\textbf{Quadratic Form and Self-Adjoint Operator}}
	In this manuscript we will apply the convention that $\cH^s(e)$ is the Sobolev space of $s$-times weakly differentiable functions on an edge $e$ whose derivatives up to order $s$ are in $L^2(e)$. The norm $\|u\|_{\cH^s}$ for $u$ in the Sobolev space $\cH^s(e)$ is equivalent to the norm $\|(I-\partial_x^2)^{s/2}u\|_{L^2(e)}$ in the Lebesgue space $L^2(e)$, e.g. see \cite{Burenkov_sobolev}. Let define underlying Hilbert space, i.e. in the terminology of Gelfand triples, the pivot space
	\begin{equation}
	\label{eq:Hilbert_space}
	\cL^2(\Gamma) := \prod_{e \in \cE} L^2(e)
	\times \prod_{e \in \cE} L^2(e)
	\times \prod_{e \in \cE} L^2(e)
	\times \prod_{e \in \cE} L^2(e)
	\times \prod_{\vt \in \cV} \bC^3(\vt)
	\times \prod_{\vt \in \cV} \bC^3(\vt)
	\end{equation}
     For any elements of the form $(\Psi,\vec g_\vt^{\hspace{0.5mm}\circ},\vec \omega_\vt^{\hspace{0.5mm} \circ}) \in \cL^2(\Gamma)$ with $\Psi = [v, w, u, \eta]^T$, inner product on $\cL^2(\Gamma)$ is defined as
	\begin{equation}
	\label{eq:innerP1}
	\begin{split}
	\langle(\Psi,\vec g_\vt^{\hspace{0.5mm}\circ},\vec \omega_\vt^{\hspace{0.5mm} \circ}),(\tilde \Psi,\vec {\tilde g}_\vt^{\hspace{0.5mm}\circ},\vec {\tilde \omega}_\vt^{\hspace{0.5mm}\circ})\rangle_{\cL^2(\Gamma)} = 
	\sum_{e \in E} \langle \Psi_e , \tilde \Psi_e \rangle_{L^2(e)} +
	\sum_{\vt \in \cV } \langle\vec g_\vt^{\hspace{0.5mm}\circ} ,\vec {\tilde g}_\vt^{\hspace{0.5mm}\circ}\rangle_{\mathfrak{m}_\vt} + \sum_{\vt \in \cV} 
	\langle \vec \omega_\vt^{\hspace{0.5mm}\circ} ,\vec {\tilde \omega}_\vt^{\hspace{0.5mm}\circ}\rangle_{\mathfrak{m}_\vt} 
	\end{split}
	\end{equation}
	where $\mathfrak{m}_\vt$ is the magnitude (generally non-zero) of concentrated mass at the vertex $\vt$. Above, the first norm on the right-hand side of \eqref{eq:innerP1} is over edges, i.e.  
	\begin{equation*}
	\begin{split}
	\langle \Psi_e , \tilde \Psi_e \rangle_{L^2(e)} := 
	\langle v_e, \tilde v_e\rangle_{L^2(e)} +
	\langle w_e, \tilde w_e\rangle_{L^2(e)} +
	\langle u_e, \tilde u_e\rangle_{L^2(e)} +
	\langle \eta_e, \tilde \eta_e\rangle_{L^2(e)}
	\end{split}
	\end{equation*}
	Additionally, denote by $\cH_\cS(\Gamma)$ to be a Hilbert space defined as
	\begin{equation}
	\label{eq:Vspace}
	\cH_\cS(\Gamma) := \prod_{e \in \cE} \cH^2(e)
	\times \prod_{e \in \cE} \cH^2(e)
	\times \prod_{e \in \cE} \cH^1(e)
	\times \prod_{e \in \cE} \cH^1(e)
	\times \prod_{\vt \in \cV} \bC^3(\vt)
	\times \prod_{\vt \in \cV} \bC^3(\vt)
	\end{equation}
	This space is equipped with the norm induced by inner product of constituent Sobolev norms, namely for elements $(\Psi,\vec g_\vt^{\hspace{0.5mm}\circ},\vec \omega_\vt^{\hspace{0.5mm} \circ}) \in \cH_\cS(\Gamma)$, then 
	\begin{equation*}
	\begin{split}
	\langle(\Psi,\vec g_\vt^{\hspace{0.5mm}\circ},\vec \omega_\vt^{\hspace{0.5mm} \circ}),(\tilde \Psi,\vec {\tilde g}_\vt^{\hspace{0.5mm}\circ},\vec {\tilde \omega}_\vt^{\hspace{0.5mm}\circ})\rangle_{\cH_\cS(\Gamma)} &= 
	\sum_{e \in \cE} \big(\langle v_e, \tilde v_e\rangle_{\cH^2(e)} +
	\langle w_e, \tilde w_e\rangle_{\cH^2(e)} +
	\langle u_e, \tilde u_e\rangle_{\cH^1(e)} +
	\langle \eta_e, \tilde \eta_e\rangle_{\cH^1(e)} \big)\\ 
	&+\sum_{\vt \in \cV } \langle\vec g_\vt^{\hspace{0.5mm}\circ} ,\vec {\tilde g}_\vt^{\hspace{0.5mm}\circ}\rangle_{\mathfrak{m}_\vt} + \sum_{\vt \in \cV} 
	\langle \vec \omega_\vt^{\hspace{0.5mm}\circ} ,\vec {\tilde \omega}_\vt^{\hspace{0.5mm}\circ}\rangle_{\mathfrak{m}_\vt} 
	\end{split}
	\end{equation*}
	Define a Sesqulinear form $\cS:\cL^2(\Gamma) \times \cL^2(\Gamma) \rightarrow \bC$ as
	\begin{equation}
	\label{eq:sesqulinearForm}
	\begin{split}
	\cS[(\Psi,\vec g_\vt^{\hspace{0.5mm}\circ},\vec \omega_\vt^{\hspace{0.5mm} \circ}),(\tilde \Psi,\vec {\tilde g}_\vt^{\hspace{0.5mm}\circ},\vec {\tilde \omega}_\vt^{\hspace{0.5mm}\circ})] :=
	\sum_{e \in \cE} \cS\up{e}[\Psi,\tilde \Psi] + 
	\sum_{v \in \cV} \cS\up{\vt}[(\vec g_\vt^{\hspace{0.5mm}\circ},\vec \omega_\vt^{\hspace{0.5mm} \circ}),(\vec {\tilde g}_\vt^{\hspace{0.5mm}\circ},\vec {\tilde \omega}_\vt^{\hspace{0.5mm}\circ})]
	\end{split}
	\end{equation}
	constructed out of a form associate to the edge set  
	\begin{equation}
	\label{eq:Seq_e}
		\cS\up{e} := \int_e \big( a_e v_e''\cc{\widetilde v_e''}
		+ b_e  w_e''\cc{\widetilde w_e''}
		+ c_e  u_e'\cc{\widetilde u_e'}
		+ d_e  \eta_e'\cc{\widetilde \eta_e'}  \big) dx 
	\end{equation}
	and vertex set of the underlying graph 
	\begin{equation}
	\label{eq:Seq_v}
	\cS\up{\vt} := \sum_{e \sim \vt} \langle \cB_e \vec g_\vt^{\hspace{0.5mm} \circ}-\vec g_{e}(\vt) ,\cB_e \vec {\tilde g}_\vt^{\hspace{0.5mm} \circ}-\vec{\tilde g}_{e}(\vt) \rangle_{\cK_{g_e}\up{\vt}} +
	\sum_{e \sim \vt} \langle \cB_e \vec \omega_\vt^{\hspace{0.5mm} \circ}-\vec \omega_{e}(\vt) ,\cB_e \vec {\tilde \omega}_\vt^{\hspace{0.5mm} \circ}-\vec{\tilde \omega}_{e}(\vt)\rangle_{\cK_{\omega_e}\up{\vt}}
	\end{equation}
	Following proposition provides a formal mathematical description of the energy form on graph with semi-rigid joint model at the vertex set. Its proof is similar to the proof of Theorem 3.1 in \cite{BE21} with slight modification and we will present it here for sake of completeness. First let us stress that Equation~\eqref{eq:Hilbert_space} specifies the underlying inner product, while \eqref{eq:Vspace} prescribes the
	correct smoothness requirements on the individual fields. The sesqulinear form \eqref{eq:sesqulinearForm} is decomposed to the (virtual) energies at edge \eqref{eq:Seq_e} and vertex \eqref{eq:Seq_v} sets, where the latter one is absent in the case of rigid-vertex assumption.  
	\begin{prop}
		\label{sesQulinearProperty}
		Energy functional \eqref{eq:energyFunctionalGraph} of the beam frame with free semi-rigid joints is the quadratic form corresponding to the positive closed sesquilinear form \eqref{eq:sesqulinearForm} densely defined on the Hilbert space $\cL^2(\Gamma)$ stated in \eqref{eq:Vspace} with domain of $\cS$ in \eqref{eq:sesqulinearForm} consisting of vectors $(\Psi,\vec g_\vt^{\hspace{0.5mm}\circ},\vec \omega_\vt^{\hspace{0.5mm} \circ}) \in \cH_\cS(\Gamma)$.  
	\end{prop}
\begin{proof}[\normalfont \textbf{Proof of Proposition~\ref{sesQulinearProperty}}] 
	For (fixed) real-valued symmetric positive definite matrices  $\{\cK_{g_e}\up{\vt}\}_{e\sim \vt}$ and $\{\cK_{\omega_e}\up{\vt}\}_{e\sim \vt}$, the sesqulinear form $\cS$ in \eqref{eq:sesqulinearForm} is obviously positive and symmetric,
	\begin{equation}
	\label{eq:positive_symm_def}
	\begin{split}
	&\cS[(\Psi,\vec g_\vt^{\hspace{0.5mm}\circ},\vec \omega_\vt^{\hspace{0.5mm} \circ}),(\Psi,\vec g_\vt^{\hspace{0.5mm}\circ},\vec \omega_\vt^{\hspace{0.5mm} \circ})] \geq 0 \\
	&\cS[(\Psi,\vec g_\vt^{\hspace{0.5mm}\circ},\vec \omega_\vt^{\hspace{0.5mm} \circ}),(\tilde \Psi,\vec {\tilde g}_\vt^{\hspace{0.5mm}\circ},\vec {\tilde \omega}_\vt^{\hspace{0.5mm}\circ})]  =
	\bar \cS[(\tilde \Psi,\vec {\tilde g}_\vt^{\hspace{0.5mm}\circ},\vec {\tilde \omega}_\vt^{\hspace{0.5mm}\circ}),(\Psi,\vec g_\vt^{\hspace{0.5mm}\circ},\vec \omega_\vt^{\hspace{0.5mm} \circ})] 
	\end{split}
	\end{equation}
	Next we establish that $\cS$ is closed, i.e. the domain $\cD(\cS)$, the subspace of $\cH_\cS(\Gamma)$, is complete with respect to the norm 
	\begin{equation}
	\label{eq:Snorm}
	\|(\Psi,\vec g_\vt^{\hspace{0.5mm}\circ},\vec \omega_\vt^{\hspace{0.5mm} \circ})\|_{\cS} := \|(\Psi,\vec g_\vt^{\hspace{0.5mm}\circ},\vec \omega_\vt^{\hspace{0.5mm} \circ})\|_{L^2(\Gamma)} + \cS[(\Psi,\vec g_\vt^{\hspace{0.5mm}\circ},\vec \omega_\vt^{\hspace{0.5mm} \circ}),(\Psi,\vec g_\vt^{\hspace{0.5mm}\circ},\vec \omega_\vt^{\hspace{0.5mm} \circ})]
	\end{equation}
	First, note that the space $\cH_\cS(\Gamma)$ is complete with respect to its norm, the sum of Sobolev norms of individual Sobolev spaces and weighted norms on $\bC^3$. Therefore $\cD(\cS)$ is a closed subspace of $\cH_\cS(\Gamma)$ and
	thus also complete with respect to the norm of  $\cH_\cS(\Gamma)$. Since for components $v$ and $w$ inequality 
	\begin{equation}
	\label{eq:itermid_deriv}
	\|f'\|_{L^2(e)}
	\leq \alpha \big(|e|^{-1} \|f\|_{L^2(e)} + |e|
	\|f''\|_{L^2(e)}\big)
	\end{equation}
	holds for some $\alpha>0$ independent of $f$, along with application of bounds on vertex values of components of $\Psi$ and their derivatives (see \cite{Burenkov98,GM20}), then
	\begin{align}
	\label{eq:Sobolev1_trace}
	|f(\vt_e)|^2
	&\leq 2|e|^{-1} \|f\|^2_{L^2(e)} + |e|\, \|f'\|^2_{L^2(e)},
	\qquad \hspace{2.8mm} \mbox{for any } f\in \cH^1(e),
	\\
	\label{eq:Sobolev2_trace}
	|f'(\vt_e)|^2
	&\leq 2|e|^{-1} \|f'\|^2_{L^2(e)} + |e|\, \|f''\|^2_{L^2(e)},
	\hspace{9.2mm} \mbox{for any } f\in \cH^2(e),
	\end{align}
	where $|e|$ is the length of the edge $e$ and $\vt_e$ is one of its endpoints. Taking into account that matrices  $\{\cK_{g_e}\up{\vt}\}_{e\sim \vt}$, $\{\cK_{\omega_e}\up{\vt}\}_{e\sim \vt}$ are elements of $S_+^3$ along with the relations for vectors $\vec g_\vt^{\hspace{0.5mm}\circ}$ and $\vec \omega_\vt^{\hspace{0.5mm} \circ}$ in terms of fields stated in \eqref{eq:semirigidVertexDefn}, then $\cS$-norm \eqref{eq:Snorm} is equivalent to the
	Sobolev norm of  $\cH_\cS(\Gamma)$. 
\end{proof}
Stem on the semi-boundedness and closeness of the form $\cS$, it corresponds to a self-adjoint differential operator or Hamiltonian on the metric graph. Main result of this paper stated in the following Theorem will characterize this differential operator and its domain.   
	\begin{thm}
		\label{MainThm}
		Energy form~\eqref{eq:energyFunctionalGraph} on a beam frame with free semi-rigid joints corresponds to the non-negative self-adjoint operator $\cA \colon \cL^2(\Gamma) \to \cL^2(\Gamma)$ with compact resolvant. Operator $\cA$ on set of edges $e \in \cE$ and vertcies $\vt \in \cV$ of the graph acting as
		\begin{equation}
			\label{diffSystem}
			\begin{pmatrix}
				\begin{pmatrix}
					v_e\\
					w_e\\
					u_e\\
					\eta_e
				\end{pmatrix}_{e \in \cE}
				, (\vec g_\vt^{\hspace{0.5mm}\circ})_{\vt \in \cV}, ( \vec \omega_\vt^{\hspace{0.5mm}\circ})_{\vt \in \cV}
			\end{pmatrix}
			\mapsto
			\begin{pmatrix}
				\begin{pmatrix}
					a_e v_e''''\\
					b_e w_e''''\\
					-c_e u_e''\\
					-d_e \eta_e''
				\end{pmatrix}_{e \in \cE}
				,\big(\mathfrak{m}_\vt^{-1} \vec F_\vt \big)_{\vt \in \cV}, ~\big(\mathfrak{m}_\vt^{-1} \vec M_\vt \big)_{\vt \in \cV}
			\end{pmatrix}
		\end{equation}
		where above $\vec F_\vt$ and $\vec M_\vt$ are respectively the net forces and moments at vertex $\vt$, i.e. 
		\begin{equation}
		\label{eq:FM}
			\begin{split}
				\vec F_\vt := \displaystyle\sum_{e \sim \vt} s_e^\vt(c_e u_e' \vec i_e - b_e w_e''' \vec j_e - a_e v_e''' \vec k_e), \qquad
				\vec M_\vt := \displaystyle\sum_{e \sim \vt} s_e^\vt (d_e \eta_e' \vec i_e - a_e v_e'' \vec j_e+  b_e w_e'' \vec k_e)
			\end{split}
		\end{equation}
		Domain of the operator $\cA$ consists of the functions $(\Psi, \vec g_\vt^{\hspace{0.5mm}\circ},\vec \omega_\vt^{\hspace{0.5mm} \circ})$ belong to 
		\begin{equation}
		\cH_\cA(\Gamma) := 
		\prod_{e \in \cE} \cH^4(e)
		\times \prod_{e \in \cE} \cH^4(e)
		\times \prod_{e \in \cE} \cH^2(e)
		\times \prod_{e \in \cE} \cH^2(e)
		\times \prod_{\vt \in \cV} \bC^3(\vt)
		\times \prod_{\vt \in \cV} \bC^3(\vt)
		\end{equation}
		that satisfy at each vertex $\vt$ and for all $e \sim \vt$, non-homogeneous Robin conditions \eqref{eq:semiRigid_displacement} and \eqref{eq:semiRigid_rotation}.
	\end{thm}	
	\begin{proof}[\normalfont \textbf{Proof of Theorem~\ref{MainThm}}] 
		Proof of the Theorem is based on extension of theoretical result in \cite{MR08} towards vector-valued Hamiltonian and adaptation of arguments in \cite{BE21} by generalized of vertex model. The reason for $\cA$ to be a self-adjoint operator with a compact resolvant, is that it is the Friedrichs extension of the triple $(\cL^2(\Gamma),\cH_\cS(\Gamma),\cS)$ which can be applied here due to the observations in  Proposition \ref{sesQulinearProperty}. Thus the operator $\cA_\cS$ is self-adjoint with domain
		\begin{equation*}
			\begin{split}
				\cD(\cA_\cS) = \Big\{(&\Psi,\vec g_\vt^{\hspace{0.5mm}\circ},\vec \omega_\vt^{\hspace{0.5mm} \circ}) \in \cH_\cS(\Gamma) ~:~ \exists  (\hat \Psi,\vec {\hat g}_\vt^{\hspace{0.5mm}\circ},\vec {\hat \omega}_\vt^{\hspace{0.5mm}\circ}) \in \cL^2(\Gamma) \text{ so that } \\
				&\cS[(\Psi,\vec g_\vt^{\hspace{0.5mm}\circ},\vec \omega_\vt^{\hspace{0.5mm} \circ}),(\tilde \Psi,\vec {\tilde g}_\vt^{\hspace{0.5mm}\circ},\vec {\tilde \omega}_\vt^{\hspace{0.5mm}\circ})] = \langle(\hat \Psi,\vec {\hat g}_\vt^{\hspace{0.5mm}\circ},\vec {\hat \omega}_\vt^{\hspace{0.5mm}\circ}),(\tilde \Psi,\vec {\tilde g}_\vt^{\hspace{0.5mm}\circ},\vec {\tilde \omega}_\vt^{\hspace{0.5mm}\circ})\rangle_{\cL^2(\Gamma)}, ~\forall (\tilde \Psi,\vec {\tilde g}_\vt^{\hspace{0.5mm}\circ},\vec {\tilde \omega}_\vt^{\hspace{0.5mm}\circ}) \in \cH_\cS(\Gamma) \Big\}
			\end{split}
		\end{equation*}  
		Thereby, two parts integration in the expression of the sesqulinear form $\cS$ leads to, $(\Psi,\vec g_\vt^{\hspace{0.5mm}\circ},\vec \omega_\vt^{\hspace{0.5mm} \circ}) \in \cD(\cA_\cS)$ if and only if $(\Psi,\vec g_\vt^{\hspace{0.5mm}\circ},\vec \omega_\vt^{\hspace{0.5mm} \circ}) \in \cH_{\cS}(\Gamma)$ and there exists $(\hat \Psi,\vec {\hat g}_\vt^{\hspace{0.5mm}\circ},\vec {\hat \omega}_\vt^{\hspace{0.5mm}\circ}) \in \cL^2(\Gamma)$ such that for any $(\tilde \Psi,\vec {\tilde g}_\vt^{\hspace{0.5mm}\circ},\vec {\tilde \omega}_\vt^{\hspace{0.5mm}\circ}) \in \cH_\cS(\Gamma)$, then  
	    \begin{equation}
	    \label{eq:allTerms}
		\begin{split}
		&\sum_{e \in \cE} \int_e \big(v_e''''\cc{\widetilde v_e}
		+ b_e  w_e''''\cc{\widetilde w_e} - c_e  u_e''\cc{\widetilde u_e}
		- d_e  \eta_e''\cc{\widetilde \eta_e} \big) dx  + 
		\sum_{\vt \in \cV} \sum_{e \sim \vt}\langle \cB_e \vec g_\vt^{\hspace{0.5mm} \circ}-\vec g_{e},\cB_e \vec {\tilde g}_\vt^{\hspace{0.5mm} \circ}-\vec{\tilde g}_{e}\rangle_{\cK_{g_e}\up{\vt}} + \\
		&\sum_{\vt \in \cV} \sum_{e \sim \vt} \langle \cB_e \vec \omega_\vt^{\hspace{0.5mm} \circ}-\vec \omega_{e},\cB_e \vec {\tilde \omega}_\vt^{\hspace{0.5mm} \circ}-\vec{\tilde \omega}_{e}\rangle_{\cK_{\omega_e}\up{\vt}} + \sum_{\vt \in V} \mathbf{B}_\vt(\Psi,\tilde \Psi) = 
		\sum_{\vt \in \cV}\langle \vec g_\vt^{\hspace{0.5mm} \circ},\vec {\tilde g}_\vt^{\hspace{0.5mm} \circ}\rangle_{\mathfrak{m}_\vt} +
		\sum_{\vt \in \cV} \langle \vec \omega_\vt^{\hspace{0.5mm} \circ},\vec {\tilde \omega}_\vt^{\hspace{0.5mm} \circ}\rangle_{\mathfrak{m}_\vt}
		\end{split}
		\end{equation}
	    Above $\mathbf{B}_\vt$ is a boundary term arises from integration by-parts, and has a form (evaluated at $\vt$)
		\begin{align}
		\label{eq:boundaryTerm}
		\mathbf{B}_\vt(\Psi,\tilde \Psi) := \sum_{e \sim \vt} (a_e v''_e\cc{\widetilde v_e'} - a_e v'''_e  \cc{\widetilde v_e} + b_e w''_e \cc{\widetilde w_e'} - b_e  w'''_e \cc{\widetilde w_e} + c_e u'_e \cc{\widetilde u_e} + d_e \eta'_e \cc{\widetilde \eta_e})
		\end{align}
		Applying the fact that set of $\{\vec i_e, \vec j_e, \vec k_e\}_{e \sim \vt}$ are orthogonal basis on edge $e$, along with realization of displacement vector $\vec g_e(\vt)$ stated in \eqref{eq:dispComp}, then 
		\begin{equation*}
			c_e u'_e \cc{\widetilde u_e} - b_e w'''_e \cc{\widetilde w_e} - a_e v'''_e  \cc{\widetilde v_e} = \langle c_e u_e' \vec i_e - b_e w_e''' \vec j_e - a_e v_e''' \vec k_e,\vec{\tilde g}_e\rangle 
		\end{equation*}
		Next, we will apply the form of force and moment stated in \eqref{eq:FeMe}. Adding and subtracting $\cB_e \vec{\tilde g}_\vt^\circ$ from the right-hand side of the above expression, summing over $e \sim \vt$, along with the fact that for each $e \sim \vt$, property $\langle \vec f_e,\cB_e \vec{\tilde g}_\vt^\circ \rangle = 
		\langle \cB_e^T \vec f_e,\vec{\tilde g}_\vt^\circ \rangle$ holds, then	
		\begin{equation}
		\label{eq:IdF}
		\begin{split}
		\sum_{e \sim \vt} (c_e u'_e \bar{\tilde u}_e - b_e w'''_e \bar{\tilde w}_e - a_e v'''_e  \bar{\tilde v}_e) = 
		\langle \vec F_\vt, \vec{\tilde g}_\vt^\circ \rangle ~-
		\sum_{e \sim \vt} \langle \vec f_e,\cB_e \vec {\tilde g}_\vt^{\hspace{0.5mm} \circ}-\vec{\tilde g}_e \rangle 
		\end{split}
		\end{equation}
		Above we use the fact that vector $\vec {\tilde g}_\vt^{\hspace{0.5mm} \circ}$ is characteristic of $\vt$ and  independent from $e \sim \vt$, along with the property of summing the vectors in first entry of right-hand side is in global-coordinate system, 
		\begin{equation}
		\sum_{e \sim \vt} \langle \cB_e^T \vec f_e,\vec {\tilde g}_\vt^{\hspace{0.5mm} \circ} \rangle = 
		\langle \sum_{e \sim \vt} \cB_e^T \vec f_e,\vec {\tilde g}_\vt^{\hspace{0.5mm} \circ}\rangle = \langle \vec F_\vt,\tilde g_\vt^\circ \rangle
		\end{equation}
		Similar steps maybe applied along with realization of rotation vector $\vec \omega_e(\vt)$ stated in \eqref{eq:omega_lin} to represent
		\begin{equation}
		\label{eq:IdM}
		\begin{split}
		\sum_{e \sim \vt} (d_e \eta'_e \bar{\tilde \eta}_e + b_e w''_e \bar{\tilde w}'_e + a_e v''_e \bar{\tilde v}'_e) = 
		\langle \vec M_\vt, \vec{\tilde \omega}_\vt^\circ \rangle ~-
		\sum_{e \sim \vt} \langle \vec m_e, \cB_e \vec {\tilde \omega}_\vt^{\hspace{0.5mm} \circ}-\vec{\tilde \omega}_{e} \rangle
		\end{split}
		\end{equation}
		This along with \eqref{eq:IdF} and \eqref{eq:IdM} will be applied in \eqref{eq:boundaryTerm} to obtain the boundary term
		\begin{equation*}
		\begin{split}
		\mathbf{B}_\vt(\Psi,\tilde \Psi) = 
		\langle \vec F_\vt, \vec{\tilde g}_\vt^\circ \rangle - 
		\sum_{e \sim \vt} \langle \vec f_e,\cB_e \vec {\tilde g}_\vt^{\hspace{0.5mm} \circ}-\vec{\tilde g}_{e} \rangle ~+ 
		\langle \vec M_\vt,  \vec{\tilde \omega}_\vt^\circ \rangle -
		\sum_{e \sim \vt} \langle  \vec m_e,\cB_e \vec {\tilde \omega}_\vt^{\hspace{0.5mm} \circ}-\vec{\tilde \omega}_{e} \rangle
		\end{split}
		\end{equation*}	
		Application of $\mathbf{B}_\vt(\Psi,\tilde \Psi)$ above along with identity $\langle\vec u,\vec v\rangle = \langle\cK^{-1}\vec u,\vec v\rangle_{\cK}$ for $\vec u, \vec v \in \bC^3$, then \eqref{eq:allTerms} reduces to
		\begin{equation*}
		\begin{split}
			&\sum_{e \in \cE} \int_e \big(a_e v_e''''\cc{\widetilde v_e}
			+ b_e  w_e''''\cc{\widetilde w_e} - c_e  u_e''\cc{\widetilde u_e}
			- d_e  \eta_e''\cc{\widetilde \eta_e}\big) dx + 
			\sum_{\vt \in \cV} \langle \vec F_\vt, \vec {\tilde g}_\vt^{\hspace{0.5mm} \circ}\rangle + \sum_{\vt \in \cV} \langle \vec M_\vt, \vec {\tilde \omega}_\vt^{\hspace{0.5mm} \circ} \rangle~ -\\
			&\sum_{\vt \in \cV} \sum_{e \sim \vt} \big(\langle (\cK_{g}\up{\vt})^{-1} \vec f_e+\vec{g}_{e} - \cB_e \vec {g}_\vt^{\hspace{0.5mm} \circ},\cB_e \vec {\tilde g}_\vt^{\hspace{0.5mm} \circ}-\vec{\tilde g}_{e} \rangle_{\cK_{g_e}\up{\vt}} + \langle (\cK_{\omega}\up{\vt})^{-1}\vec m_e+\vec{\omega}_{e} - \cB_e \vec {\omega}_\vt^{\hspace{0.5mm} \circ},\cB_e \vec {\tilde \omega}_\vt^{\hspace{0.5mm} \circ}-\vec{\tilde \omega}_{e} \rangle_{\cK_{\omega_e}\up{\vt}}\big) = \\ 
			&\sum_{e \in \cE} \int_e \big(v_e\cc{\widetilde v_e}
			+ w_e\cc{\widetilde w_e} - u_e\cc{\widetilde u_e}
			- \eta_e\cc{\widetilde \eta_e}\big) dx +
			\sum_{\vt \in \cV} \langle \vec {g}_\vt^{\hspace{0.5mm} \circ}, \vec {\tilde g}_\vt^{\hspace{0.5mm} \circ} \rangle_{\mathfrak{m}_\vt} + 
			\sum_{\vt \in \cV} \langle \vec {\omega}_\vt^{\hspace{0.5mm} \circ},\vec {\tilde \omega}_\vt^{\hspace{0.5mm} \circ} \rangle_{\mathfrak{m}_\vt}
		\end{split}
		\end{equation*}
	    But, by the Definition \ref{def:semiRigid}, properties $(\cK_{g}\up{\vt})^{-1} \vec f_e+\vec{g}_{e}-\cB_e \vec {g}_\vt^{\hspace{0.5mm} \circ} = 0$ and $(\cK_{\omega}\up{\vt})^{-1}\vec m_e+\vec{\omega}_{e}-\cB_e \vec {\omega}_\vt^{\hspace{0.5mm} \circ} = 0$ hold. Moreover, applying $\langle \vec F_\vt, \vec {\tilde g}_\vt^{\hspace{0.5mm} \circ} \rangle = \langle \mathfrak{m}_\vt^{-1} \vec F_\vt, \vec {\tilde g}_\vt^{\hspace{0.5mm} \circ} \rangle_{\mathfrak{m}_\vt}$ and $\langle \vec M_\vt, \vec {\tilde \omega}_\vt^{\hspace{0.5mm} \circ}\rangle = 
	    \langle \mathfrak{m}_\vt^{-1} \vec M_\vt, \vec {\tilde \omega}_\vt^{\hspace{0.5mm} \circ} \rangle_{\mathfrak{m}_\vt}$, then \eqref{eq:allTerms} reduces to
	    \begin{equation*}
	    \begin{split}
	    &\sum_{e \in \cE} \int_e \big(a_e v_e''''\cc{\widetilde v_e}
	    + b_e  w_e''''\cc{\widetilde w_e} - c_e  u_e''\cc{\widetilde u_e}
	    - d_e  \eta_e''\cc{\widetilde \eta_e}\big) dx + 
	    \sum_{\vt \in \cV} \langle \mathfrak{m}_\vt^{-1} \vec F_\vt, \vec {\tilde g}_\vt^{\hspace{0.5mm} \circ} \rangle_{\mathfrak{m}_\vt} +
	    \sum_{\vt \in \cV} \langle \mathfrak{m}_\vt^{-1} \vec M_\vt, \vec {\tilde \omega}_\vt^{\hspace{0.5mm} \circ} \rangle_{\mathfrak{m}_\vt} = \\
	    &\sum_{e \in E} \int_e \big(v_e\cc{\widetilde v_e}
	    + w_e\cc{\widetilde w_e} - u_e\cc{\widetilde u_e}
	    - \eta_e\cc{\widetilde \eta_e}\big) dx +
	    \sum_{\vt \in \cV} \langle \vec {g}_\vt^{\hspace{0.5mm} \circ}, \vec {\tilde g}_\vt^{\hspace{0.5mm} \circ}\rangle_{\mathfrak{m}_\vt} + 
	    \sum_{\vt \in \cV} \langle \vec {\omega}_\vt^{\hspace{0.5mm} \circ}, \vec {\tilde \omega}_\vt^{\hspace{0.5mm} \circ} \rangle_{\mathfrak{m}_\vt}	
	    \end{split}
	    \end{equation*}
	    This then implies that the expression for the operator $\cA_\cS$ which coincides with that of $\cA$. Both domains also coincide. Finally, the equivalence of norms between $\|\cdot\|_{\cS}$ and $\cH_{\cS}(\Gamma)$ stated in Proposition \ref{sesQulinearProperty} proves the positiveness of operator $\cA$. 
	\end{proof}
\begin{remark}
	We stress that the net forces and moments at vertex $\vt$ in the statement of Theorem \ref{MainThm} are respectively of the form $\vec F_\vt := \sum_{e \sim \vt} \vec f_e$ and $\vec M_\vt :=  \sum_{e \sim \vt} \vec m_e$ constructed out of edgewise counterparts defined in \eqref{eq:FeMe}. These two quantities encapsulate important physical meaning, namely, they respectively model dynamics of net forces and moments developed at a vertex with concentrated mass and due to a relative displacement and torsion of edges adjacent to the vertex. This generalizes the result in \cite{BE21} in which for the case of massless rigid-joint, the two conditions are vanishing. But presence of concentrated mass causes apprentice of spectral dependent vertex conditions.   
\end{remark}
	\subsubsection{\textbf{Planar Frame}}
	A particularly well-studied case in the literature is the planar
	frame with rigid joints, which in its undeformed configuration is embedded in a two dimensional plane. It has been shown that in this situation the Hamiltonian of the frame decouples into two operators, one linking out-of-plane with angular displacements and the other one linking in-plane with axial displacements, see Corollary 4.2 in \cite{BE21} for detail. Next, we will discuss role of semi-rigid joint on validity of a similar decoupling property. Without loss of generality assume that this plane has normal vector $\vec E_3$.  We also assume that for all edges $e$, vector $\vec k_e$ is chosen to be the same as the global vector $\vec E_3$. Following definition will characterize class of (vertex) stiffness matrices which preserves decoupling of modes similar to rigid-joint case. 
	In this section for simplicity we assume global basis are the standard ones, i.e. $\vec E_i$ has only non-zero entry at position $i$. 
	\begin{defn}
		\label{kPeresev}
		Matrix $\cK \in S_+^3$ is called $\vec k$-plane preserving if it has a form
		\begin{equation}
		\label{eq:kPerserve}
			\cK := 
			\begin{pmatrix}
			\kappa_{1} & \kappa & 0 \\
			\kappa & \kappa_{2} & 0 \\
			0 & 0 & \kappa_{3}
			\end{pmatrix}
		\end{equation} 
	\end{defn}
    Let denote by $d_\kappa := \kappa_{1}\kappa_{2} - \kappa^2$, then observe that for any vector $\vec x = x_1 \vec i_e + x_2 \vec j_e + x_3 \vec k_e$ and $\cK_e$ of the form \eqref{eq:kPerserve}, then
    \begin{equation}
    \label{eq:kInvX}
    	\cK_e^{-1} \vec x = 
    	d_\kappa^{-1}(\kappa_{2} x_1-\kappa x_2) \vec i_e + 
    	d_\kappa^{-1}(\kappa_{1} x_2-\kappa x_1) \vec j_e + 
    	\kappa_3^{-1} x_3 \vec k_e
    \end{equation}
    suppresses a presence of $x_1$, and $x_2$ in the $\vec k_e$ direction. We stress that $\det(\cK) = \kappa_3d_\kappa \in \bR_+$ and applying property $\kappa_3 > 0$, thereby $d_\kappa \in \bR_+$. Special member of $\vec k$-plane preserving matrices are the diagonal ones with positive entries. In this case, \eqref{eq:kInvX} reduces to  
    	\begin{equation}
    	\cK_e^{-1} \vec x = 
    	\kappa_1^{-1} x_1 \vec i_e + 
    	\kappa_2^{-1} x_2 \vec j_e + 
    	\kappa_3^{-1} x_3 \vec k_e
    	\end{equation}
    Following corollary shows that under class of $\vec k$-plane preserving stiffness matrices, the operator decomposes into a direct sum of two operators, one coupling out-of-plane to angular displacements and the other coupling in-plane with axial displacements. Observe that for planar graphs and the convention discussed above, vectors $\vec i_e$ and $\vec j_e$ can be written of size two. Prior to state the desired result, let denote by $\cP := (\vec E_1, \vec E_2)$ to be $3\times 2$ matrix, and moreover let $\cQ := \vec E_3$.
	\begin{cor}
		\label{decouplingPlane}
		Free planar network of beams with semi-rigid joints equipped with family of $\vec k$-preserving stiffness matrices $\{\cK_{g_e}\up{\vt}\}_{e\sim \vt}$ and $\{\cK_{\omega_e}\up{\vt}\}_{e\sim \vt}$ is described by Hamiltonian $\cA = \cA\up{\text{out}} \oplus \cA\up{\text{in}}$ where $\cA\up{\text{out}}$ is a differential operator acting as
		\begin{equation}
			\label{eq:outPlane}
			\begin{Bmatrix}
				\begin{pmatrix}
					v_e\\
					\eta_e
				\end{pmatrix}_{e \in \cE}
				, (v_\vt^\circ)_{\vt \in \cV}, (\vec \omega_\vt^{\hspace{0.5mm}\circ})_{\vt \in \cV}
			\end{Bmatrix}
			\mapsto
			\begin{Bmatrix}
				\begin{pmatrix}
					a_e v_e''''\\
					-d_e \eta_e''
				\end{pmatrix}
				, \big(\mathfrak{m}_\vt^{-1} F_\vt\up{\text{out}} \big)_{\vt \in \cV}, ~\big(\mathfrak{m}_\vt^{-1} \vec M_\vt\up{\text{out}} \big)_{\vt \in \cV}
			\end{Bmatrix}
		\end{equation}
		on functions in the space
		$\prod_{e \in \cE} \cH^4(e) \times \prod_{e \in \cE} \cH^2(e) \times \prod_{\vt \in \cV} \bC(\vt) \times \prod_{\vt \in \cV} \bC^2(\vt)$ satisfying at each vertex $\vt \in \cV$ and for all $e \sim \vt$ the primary conditions
		\begin{subequations}
			\label{vertexCondPlanarPrimaryH1}
			\begin{gather}
				\label{primaryVertexCond1H1}
				v_e(\vt) - s_e^\vt \big(\cQ^T\cK_{g_e}\up{\vt}\cQ \big)^{-1}a_e v_e'''  = v_\vt^\circ \\
				\label{primaryVertexCond2H1}
				\hspace{3mm}\eta_e(\vt) \vec i_e - v_e'(\vt) \vec j_e + s_e^\vt \big(\cP^T \cK_{\omega_e}\up{\vt}\cP \big)^{-1}\big(d_e \eta_e' \vec i_e - a_e v_e'' \vec j_e\big) = \vec \omega_\vt^{\hspace{0.5mm}\circ} 
			\end{gather}
		\end{subequations}
		with net forces and moments respectively of a form
		\begin{subequations}
			\begin{gather}
				F_\vt\up{\text{out}} := -\displaystyle\sum_{e \sim \vt} s_e^\vt a_e v_e''', \qquad 
				\vec M_\vt\up{\text{out}} := \displaystyle\sum_{e \sim \vt} s_e^\vt (d_e \eta_e' \vec i_e - a_e v_e'' \vec j_e)
			\end{gather}
		\end{subequations}
		The operator $\cA\up{\text{in}}$ acts as 
		\begin{equation}
			\label{eq:inPlane}
			\begin{Bmatrix}
				\begin{pmatrix}
					w_e\\
					u_e
				\end{pmatrix}_{e \in \cE}
				,( \vec g_\vt^{\hspace{0.5mm}\circ})_{\vt \in V}, ( \omega_\vt^\circ)_{\vt \in \cV}
			\end{Bmatrix}
			\mapsto
			\begin{Bmatrix}
				\begin{pmatrix}
					b_e w_e''''\\
					-c_e u_e''
				\end{pmatrix}
				, \big(m_\vt^{-1} \vec F_\vt\up{\text{in}} \big)_{\vt \in \cV}, ~\big(m_\vt^{-1} M_\vt\up{\text{in}} \big)_{\vt \in \cV}
			\end{Bmatrix}
		\end{equation}
		on functions in the space
		$\prod_{e \in \cE} \cH^4(e) \times \prod_{e \in \cE} \cH^2(e) \times \prod_{\vt \in \cV} \bC^2(\vt) \times \prod_{\vt \in \cV} \bC(\vt)$ satisfying at each vertex $\vt$ and for all $e \sim \vt$ the primary conditions
		conditions
		\begin{subequations}
			\label{vertexCondPlanarPrimaryH2}
			\begin{gather}
				\label{primaryVertexCond1H2}
				\hspace{3mm}u_e(\vt) \vec i_e + w_e(\vt) \vec j_e + s_e^\vt\big(\cP^T \cK_{g_e}\up{\vt}\cP\big)^{-1}\big(c_e u_e' \vec i_e - b_e w_e''' \vec j_e\big) = \vec g_\vt^{\hspace{0.5mm}\circ} \\
				\label{primaryVertexCond2H2}
				w_e'(\vt) + s_e^\vt \big(\cQ^T\cK_{\omega_e}\up{\vt}\cQ \big)^{-1} b_e w_e'' = \omega_\vt^\circ
			\end{gather}
		\end{subequations}
		with net forces and moments respectively of a form
		\begin{subequations}
			\begin{gather}
				\vec F_\vt\up{\text{in}} :=\displaystyle\sum_{e \sim \vt} s_e^\vt(c_e u_e' \vec i_e - b_e w_e''' \vec j_e), \qquad
				M_\vt\up{\text{in}} := \displaystyle \sum_{e \sim \vt}  s_e^\vt b_e w_e'' 
			\end{gather}
		\end{subequations}
	\end{cor}
	\begin{proof}[\normalfont \textbf{Proof of Corollary~\ref{decouplingPlane}}] 
	The differential expression for the operator $\cA$ is already in the ``block-diagonal'' form, see \eqref{diffSystem}, so it remains to show that the vertex conditions decompose as described. But that follows directly from projecting conditions \eqref{eq:semiRigid_displacement}, \eqref{eq:semiRigid_rotation} and later \eqref{eq:S} onto the common normal $\vec k$ and onto its orthogonal complement along with application of definition \eqref{kPeresev}. 
	\end{proof}
	\section{Eigenvalue problem on Compact Graph}
	\label{sec:CompactGraph}
	\subsection{Preliminary}
	Our aim in the remaining Sections is to characterize the spectrum $\sigma(\cA)$ of operator $\cA$ acting on compact graphs \footnote{In our setting compact frames means: that $\cE$ is a finite set and each edge $e \in \cE$ has finite length.}. According to the Proposition \ref{sesQulinearProperty} this spectrum is positive and discrete. As it is common in literature, we shall rewrite the eigenvalue problem into an equivalent matrix differential value problem, known as  ``characteristic'' or ``secular'' equation \cite{Banerjee19, KotSmi_ap99}. The most direct approach is to solve the eigenvalue equation $\cA \Psi = \lambda \Psi$ component-wise on every edge before applying conditions at vertices. We will show detail of this derivation for a simple 1D example (known as Cantilevered beam) in which discontinuity of fields are admissible. However, problems can quickly become computationally overwhelming due to large number of degrees of freedom associated to each beam. Techniques based on representation theory of finite groups maybe applied on a highly symmetric graph example to decompose the original problem into a sum of operators each corresponding to a particular class of vibrational modes, e.g. see \cite{BS04, BE21}. We will discuss extension of this result for the case in which concentrated mass exists at the semi-rigid joint. Another approach to deal with more general problems (e.g. absence of symmetry) is to employ numerical schemes such as finite element method, e.g. see \cite{AB17}. The price for that however, is losing the analytic form of secular form and requirements of dense frame's discretization for capturing high energy modes. Stem on the techniques applied in this framework, we will discuss the idea of constructing geometric-free local spectral basis set together with enforcing geometry of compact graph into play by representation the global solution as a linear combination of them. This will be the topic of next Section and a departure point for the forthcoming work on spectral analysis of periodic frames equipped with general class of vertex matching models \cite{ES21}.    
	
	Here, we recall that for the general semi-rigid vertex model, at each $\vt \in \cV$ and for all $e \sim \vt$, conditions \eqref{eq:semirigidVertexDefn} should be enforced. Moreover, presence of concentrated mass ($\mathfrak{m}_\vt > 0$) will turn the net force and moment conditions depending on the eigenvalue, namely at each  $\vt \in \cV$ properties 
	\begin{subequations}
		\label{eq:S}
		\begin{gather}
			\label{eq:S1}
			\sum_{e \sim \vt} s_e^\vt(c_e u_e' \vec i_e - b_e w_e''' \vec j_e
			- a_e v_e''' \vec k_e) = \lambda \mathfrak{m}_\vt  \vec g_\vt^{\hspace{0.5mm} \circ}, \\
			\label{eq:S2}
			\sum_{e \sim \vt} s_e^\vt(d_e \eta_e' \vec i_e - a_e v_e'' \vec j_e
			+  b_e w_e'' \vec k_e) = \lambda \mathfrak{m}_\vt \vec \omega_\vt^{\hspace{0.5mm} \circ}.
		\end{gather}
	\end{subequations}
	are imposed. Finally, it maybe interesting to observe that vertex conditions \eqref{eq:semiRigid_displacement} and \eqref{eq:S1} can be represented in a well-known compact format established for class of second-order {S}chr{\"o}dinger operators, e.g. see \cite{BK13}. For $e\sim \vt$, let $\vec g_e$ and $\vec f_e$ be vectors of size 3 with entries of displacement and force vectors corresponding $e_\vt$ in local coordinate of $e$. Similar definition holds for vectors $\vec \omega_e$ and $\vec m_e$. Introduce matrices 
	\begin{myequation}
		\mathbb{A}_\vt :=
		\begin{pmatrix}
		\cB_1 & -\cB_2 & \hspace{2mm}0 & \cdots & 0 & 0 \\
		0     & \hspace{3mm}\cB_2 & -\cB_3 & \cdots &0 & 0 \\
		\cdots & \cdots & \cdots & \cdots &\cdots & \cdots \\
		0 & 0 & 0 & \cdots & \cB_{n-1} & -\cB_n \\
		-\lambda \mathfrak{m}_\vt \cB_1 & 0 & 0 & \cdots & 0 & 0
		\end{pmatrix}, \quad
		\mathbb{B}_\vt := 
		\begin{pmatrix}
		s_1\cK_{1}^{-1} & -s_2\cK_{2}^{-1} & 0 &\cdots & 0 & 0 \\
		0  & \hspace{3mm}s_2\cK_{2}^{-1}& -s_3\cK_{3}^{-1} & \cdots &0 & 0\\
		\cdots  & \cdots & \cdots & \cdots & \cdots & \cdots\\
		0 &  0 & 0 & \cdots & s_{n-1}\cK_{n-1}^{-1} & -s_n\cK_{n}^{-1} \\
		s_1\cB_1 & s_2\cB_2  & s_3\cB_3 &\cdots & s_{n-1}\cB_{n-1} & s_n\cB_n
		\end{pmatrix}
	\end{myequation}
	where $\cB_i$ is the transformation matrix of an edge $e_i$ introduced in Section \ref{sec:Elasticity} and $0$ is zero-block matrix of size $3\times 3$. Moreover, let denote by $\mathbf{g}\up{\vt} := [\vec g_1~\vec g_2~\cdots \vec g_n]^{\text{T}}$ and $\mathbf{f}\up{\vt} := [\vec f_1~\vec f_2~\cdots \vec f_n]^{\text{T}}$ to be vectors of size $(3\text{d}_\vt) \times 1$, with $\text{d}_\vt$ stands for degree of vertex $\vt$. Then conditions \eqref{eq:semiRigid_displacement} and \eqref{eq:S1} can be recast as a single equation $\mathbb{A}_\vt \mathbf{g}\up{\vt} + \mathbb{B}_\vt \mathbf{f}\up{\vt} = \mathbf{0}$. Similar result should hold for rotation and moments at $\vt$ by relation $\mathbb{A}_\vt \mathbf{\omega}\up{\vt} + \mathbb{B}_\vt \mathbf{m}\up{\vt} = \mathbf{0}$. These two systems can be then merged as
	\begin{equation}
	\label{eq:bcBlock1}
		\begin{pmatrix}
		\mathbb{A}_\vt & \mathbf{0} \\
		\mathbf{0} & \mathbb{A}_\vt
		\end{pmatrix}
		\begin{pmatrix}
		\mathbf{g}\up{\vt} \\
		\mathbf{\omega}\up{\vt}
		\end{pmatrix}
		+
        \begin{pmatrix}
		\mathbb{B}_\vt & \mathbf{0} \\
		\mathbf{0} & \mathbb{B}_\vt
		\end{pmatrix}
		\begin{pmatrix}
		\mathbf{f}\up{\vt} \\
		\mathbf{m}\up{\vt}
		\end{pmatrix}
		= 
		\begin{pmatrix}
		\mathbf{0}\\
		\mathbf{0}
		\end{pmatrix}
	\end{equation}
	which is in the form of a well-known general
	(homogeneous) vertex condition. Observe that $\mathbb{A}_\vt$ and $\mathbb{B}_\vt$ are $(3\text{d}_\vt)\times(3\text{d}_\vt)$ matrices. Due to the (Block) diagonal structure of \eqref{eq:bcBlock1}, in order to guarantee that the correct number of independent conditions (equal to $6\text{d}_\vt$) is imposed, rank of the $3\text{d}_\vt\times 6\text{d}_\vt$ global pair matrices $(\mathbb{A}_\vt,\mathbb{B}_\vt)$ must be equal to $3\text{d}_\vt$, i.e., maximal. It is easy to check that this property holds for the proposed general class of semi-rigid joint model and the case in which concentrated mass exists.   
	\subsection{Example of One Dimensional Graph}
	\label{subs:exam1}
\begin{figure}[ht]
		\centering
		\includegraphics[width=0.385\textwidth]{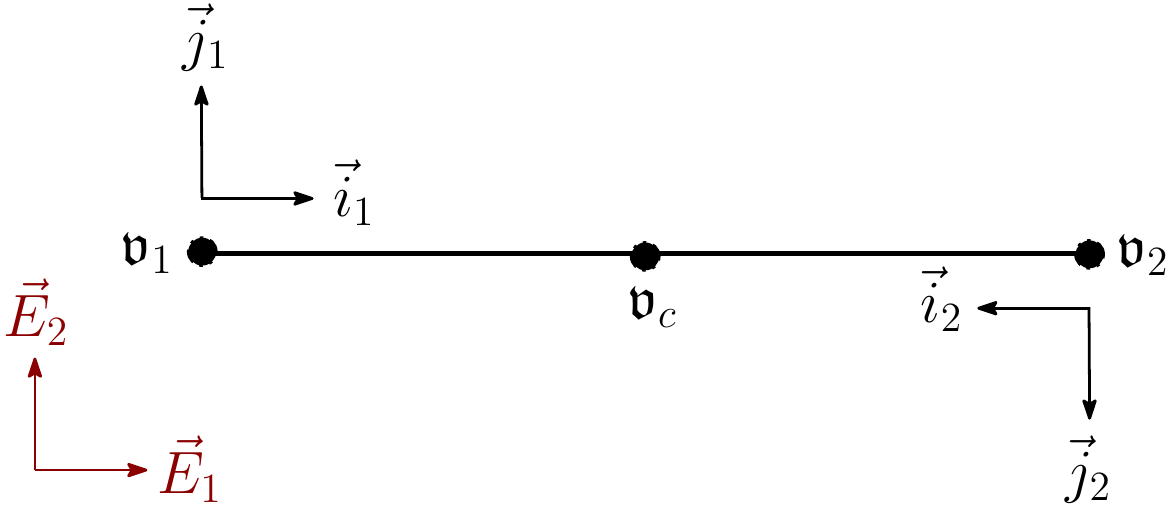}
		\caption{Geometry of 1D graph in its equilibrium with (generally) semi-rigid central vertex $\vt_c$.} 
		\label{fig:cantileverBeam}
	\end{figure}

	Consider beam frame depicted in Figure \ref{fig:cantileverBeam} consisting of two edges $e_1, e_2$ meeting at the central free semi-rigid joint $\vt_c$. The beams are oriented from the beam's ends to the joint, see the local basis of each beam. In the case of rigid central vertex, and by merging the two edges to one edge, see Remark 4.4 in \cite{BE21}, this frame is known as \emph{cantilevered} beam and all fields will be decoupled. However, this decoupling for semi-rigid vertex is only the case if stiffness matrix in \eqref{eq:S} is diagonal. For simplicity and comparison among the two types of join models, here we assume diagonal vertex stiffness and will investigate the role of different parameters on eigenfunctions corresponding to the in-plane lateral displacement field $w(x)$. General eigenvalue problem for $e =1,2$ then has a form
	\begin{equation}
	\label{eq:GW}
		b_e w_e''''(x) = \lambda w_e(x)
	\end{equation}
	subjected to fixed boundary condition at $\vt_1$, i.e. $w_1(0) = w_1'(0) = 0$ and free type at $\vt_2$, namely $w_2''(0) = w_2'''(0) = 0$, see Table 1 in \cite{BE21} for general boundary types. Applying properties $\vec j_2 = -\vec j_1$ and $\vec k_2 = \vec k_1$ along with \eqref{eq:semirigidVertexDefn} and \eqref{eq:S}, at the central vertex $\vt_c$ displacement (and rotation) relations 
	\begin{subequations}
		\label{eq:dispRotExm1}
	\begin{equation}
	w_1(\ell_1) - \kappa_{g_1}^{-1} b_1 w_1'''(\ell_1) = -w_2(\ell_2) + \kappa_{g_2}^{-1} b_2 w_2'''(\ell_2) = w_{\vt_c}^\circ
	\end{equation}
	\begin{equation}
	w_1'(\ell_1) + \kappa_{\omega_1}^{-1} b_1 w_1''(\ell_1) = + w_2'(\ell_2) + \kappa_{\omega_2}^{-1} b_2 w_2''(\ell_2) = \omega_{\vt_c}^\circ
	\end{equation}
	\end{subequations}
	along with force (and moment) constraints 
	\begin{subequations}
		\label{eq:forceMomExm1}
	\begin{equation}
	-b_1 w_1'''(\ell_1) + b_2 w_2'''(\ell_2) = \lambda \mathfrak{m}_{\vt_c} w_{\vt_c}^\circ
	\end{equation}
	\begin{equation}
	+b_1 w_1''(\ell_1) + b_2 w_2''(\ell_2) = \lambda \mathfrak{m}_{\vt_c} \omega_{\vt_c}^\circ
	\end{equation}
	\end{subequations}
	should be satisfied. In the rigid-vertex case, i.e. letting $\kappa_{g_i}^{-1}, \kappa_{\omega_i}^{-1} \rightarrow 0$, and by setting parameters $b_i = 1$, then eigenvalues $\lambda_1 \leq \lambda_2 \leq \cdots$ satisfy characteristic equation \footnote{For example see \href{https://en.wikipedia.org/wiki/Euler-Bernoulli_beam_theory}{wikipedia page}.}
	\begin{equation}
	\label{eq:1DSeq}
	\cosh(\mu_n L) \cos(\mu_n L) + 1 = 0
	\end{equation}
	with $L := \ell_1 + \ell_2$ and $\mu_n := \lambda_n^{1/4}$. Corresponding normalized eigenfunctions then have a form
	\begin{equation}
	\label{eq:cantE}
	w\up{n}(x) = A \Big[\cosh(\mu_n x) - \cos(\mu_n x) - \frac{\cosh(\mu_n L) + \cos(\mu_n L)}{\sinh(\mu_n L) + \sin(\mu_n L)}\big(\sinh(\mu_n x) - \sin(\mu_n x)\big)\Big]
	\end{equation} 

\begin{figure}[ht]
		\centering
		\includegraphics[width=1\textwidth]{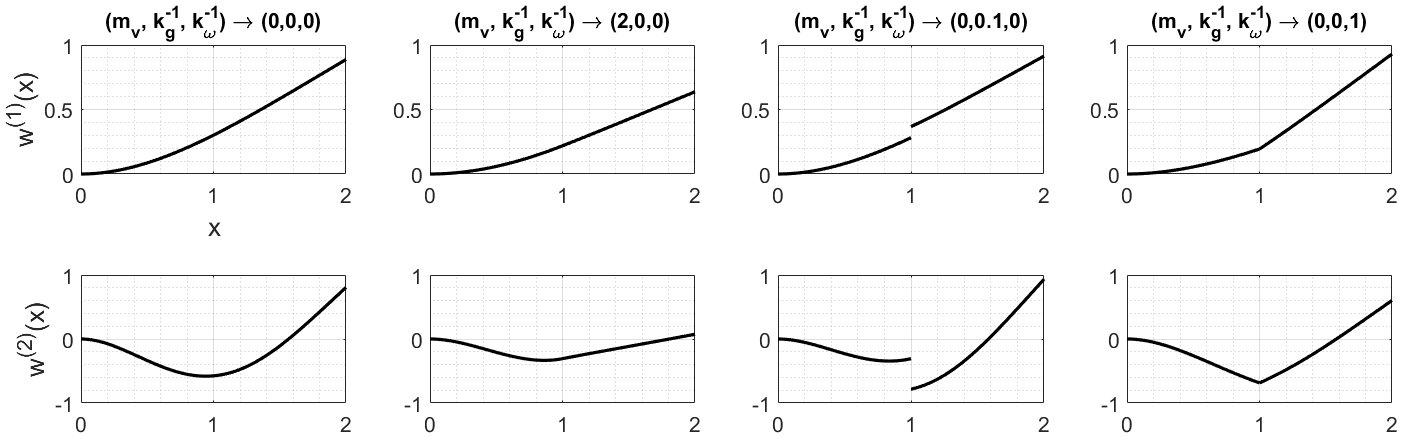}
		\caption{Plots of first two eigenfunctions $w\up{1}(x)$ and $w\up{2}(x)$ in global coordinate for different values of concentrated mass $\mathfrak{m}_{\vt_c}$, and vertex stiffness including $\kappa_{g}^{-1}$ and $\kappa_{\omega}^{-1}$.}
		\label{fig:1DExmp}
	\end{figure}

    For the semi-rigid central joint model, imposing the boundary conditions at $\vt_1, \vt_2$, then general solution to the eigenvalue problem \eqref{eq:GW} on each edge reduces to 
	\begin{subequations}
		\label{eq:genSolExm1}
	\begin{equation}
	\label{eigenProblemPlanar1}
	w_1\up{n}(x) = A_1\up{n} (\sinh(\mu_n x) - \sin(\mu_n x))
	+ B_1\up{n} (\cosh(\mu_n x) - \cos(\mu_n x)),
	\end{equation}
	\begin{equation}
	\label{eigenProblemPlanar2}
	w_2\up{n}(x) = A_2\up{n} (\sinh(\mu_n x) + \sin(\mu_n x)) + B_2\up{n} (\cosh(\mu_n x) + \cos(\mu_n x)).
	\end{equation}
	\end{subequations}
	Applying vertex conditions \eqref{eq:dispRotExm1} and \eqref{eq:forceMomExm1} in general solutions \eqref{eq:genSolExm1}, eigenvalue problem reduces to finding the ($n$-dependent) coefficient vector
	$[A_1, B_1, A_2, B_2, w_{\vt_c}^\circ, \omega_{\vt_c}^\circ]^{\text{T}}$ in the kernel of the $6\times6$ matrix $\mathbb{M}(\lambda)$ given by
	\begin{myequation1}
	\begin{pmatrix}
	\hspace{2.5mm}S^-_{\mu \ell_1} - \mu^3 \kappa_{g_1}^{-1} C^+_{\mu \ell_1} &\hspace{1mm} C^-_{\mu \ell_1} - \mu^3  \kappa_{g_1}^{-1} S^-_{\mu \ell_1} & 
	\hspace{1mm}0 & 0 & -1& \hspace{6mm}0\\
	C^-_{\mu \ell_1} + \mu \kappa_{\omega_1}^{-1} S^+_{\mu \ell_1} &S^+_{\mu \ell_1}+ \mu \kappa_{\omega_1}^{-1} C^+_{\mu \ell_1} & \hspace{1mm}0 & 0 &\hspace{3mm}  0& \hspace{3mm} -1 \\
	0 &0 &\hspace{6mm} S^+_{\mu \ell_2} - \mu^3 \kappa_{g_2}^{-1} C^-_{\mu \ell_2} &\hspace{5mm}
	C^+_{\mu \ell_2} - \mu^3  \kappa_{g_2}^{-1} S^+_{\mu \ell_2}& \hspace{3mm}1& \hspace{6mm}0\\
	0 &0&\hspace{3.25mm} C^+_{\mu \ell_2} + \mu \kappa_{\omega_2}^{-1} S^-_{\mu \ell_2} &\hspace{2.5mm} S^-_{\mu \ell_2} + \mu \kappa_{\omega_2}^{-1} C^-_{\mu \ell_2} &\hspace{3mm}  0& \hspace{3mm} -1 \\
	C^+_{\mu \ell_1} & \hspace{2mm}S^-_{\mu \ell_1} &\hspace{2mm} 
	-C^-_{\mu \ell_2} &\hspace{1.2mm} -S^+_{\mu \ell_2} &\hspace{4mm} \mu \mathfrak{m}_{\vt_c} &\hspace{6mm} 0 \\
	S^+_{\mu \ell_1} &\hspace{2mm} C^+_{\mu \ell_1} & 
	\hspace{4mm}S^-_{\mu \ell_2} &\hspace{4mm} C^-_{\mu \ell_2} &\hspace{3mm} 0 & \hspace{3mm}-\mu^2 \mathfrak{m}_{\vt_c}
	\end{pmatrix}
	\end{myequation1}
	with the corresponding eigenvalue $\lambda_n$ being a solution of \eqref{eq:GW}. Above (and in the rest of paper) we used the following abbreviations to make the matrix presentation more compact
	\begin{equation}
		\label{eq:notationMatrix}
		S_{\gamma} := \sin(\gamma), \hspace{5mm}  
		C_{\gamma} := \cos(\gamma), \hspace{5mm}
		S_{\gamma}^{\pm} := \sinh(\gamma) \pm \sin(\gamma), \hspace{5mm} 	C_{\gamma}^{\pm} := \cosh(\gamma) \pm \cos(\gamma),
	\end{equation}
	Figure \ref{fig:1DExmp} plots eigenfunctions corresponding to the first two eigenvalues of $\mathbb{M}$ for different set of parameters. Compare to the cantilevered beam \eqref{eq:cantE}, it is clear that how introducing semi-rigid central vertex will cause discontinuity of $w(x)$ and its derivative. 

     Referring to the simple example above, it is clear that deriving analytical solutions for more complicated graph in which all degrees of freedom active is not a desirable (and maybe computationally feasible) task. In the work \cite{BE21}, application of group representation theory to classify eigenmodes and simplifying analysis of graphs with a high degree of symmetry for class of rigid-vertex model has been discussed. Main goal of next example is to extend this result to semi-rigid joint model equipped with concentrated mass at a vertex. 
	\subsection{Example of Three Dimensional Graph.}
	\label{sec:example_antenna}
\begin{figure}[ht]
    	\centering
    	\includegraphics[width=0.35\textwidth]{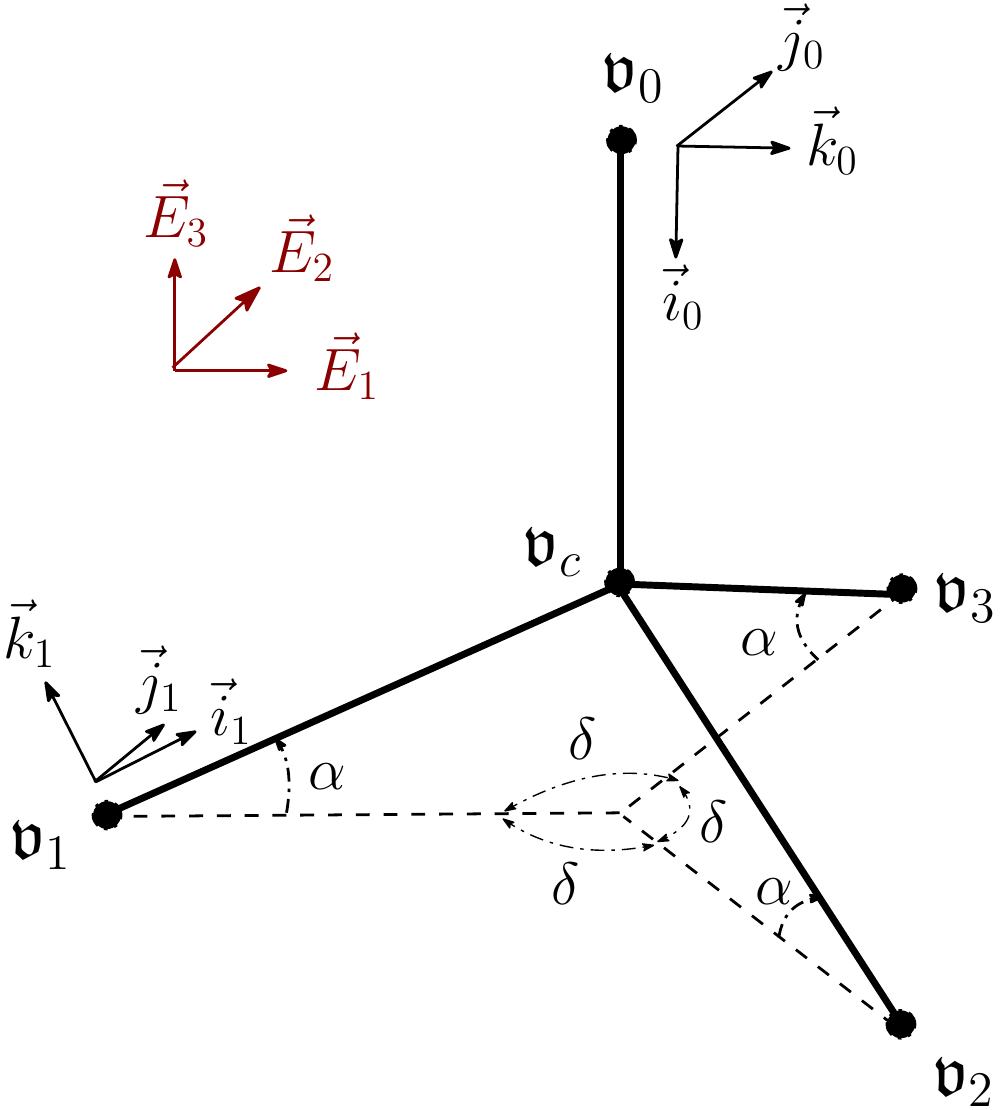}
    	\caption{Geometry of three dimensional star graph in its equilibrium sate with (generally) semi-rigid central vertex $\vt_c$. Variation of angle $\alpha$ generates various geometries but all satisfy result of Theorem \ref{thm:reducing_antenna_tower}.}
    	\label{threeDimGraphExample}
    \end{figure}

	Let $\Gamma$ be the graph formed by three leg beams
	$e_1, e_2, e_3$ and a vertical beam $e_0$, all joining at the central vertex $\vt_c$, see Figure~\ref{threeDimGraphExample}. The structure and its material parameters are assumed to be symmetric with respect to rotation by $2\pi/3$ around the vertical axis and with respect to reflection swapping a pair of the leg beams, i.e. 
	\begin{assum}
		\label{symAntenaAssum}
		The symmetric graph $\Gamma$ satisfies
	\begin{itemize}
		\item[(i)] Beam's material parameters satisfy $a_0=b_0$ and that the leg beams are identical with principal axes of inertia that may be chosen to have $\vec{j}_s \perp \vec{E}_3$
		for $s=1,2,3$.
		\item[(ii)] Displacement and rotation stiffness corresponding semi-rigid central vertex $\vt_c$, for vertical edge is a pair $(\kappa_{g_0},\kappa_{\omega_0})$ and for the leg edges satisfy $(\kappa_{g_s},\kappa_{\omega_s}) = (\kappa_g, \kappa_\omega)$ for $s=1,2,3$.
	\end{itemize}
	\end{assum}
	The leg ends $\vt_1,\vt_2,\vt_3$ are fixed (vanishing displacement and angular displacement), vertex $\vt_c$ is a free semi-rigid joint and the end $\vt_0$ of the vertical beam is free. Our main result of this Section is a decomposition of the operator $\cA$ into a direct sum of four self-adjoint operators. 
	\begin{thm}[Domain Decomposition] 
		\label{thm:reducing_antenna_tower}
		The Hamiltonian operator $\cA$ of the beam frame $\Gamma$ is reduced by the decomposition
		\begin{equation}
			\label{eq:reducing_AT}
			\cL^2(\Gamma) = \cH_{\mathrm{id}} \times \cH_{\mathrm{alt}}
			\times \cH_\omega \times \cH_{\cc{\omega}},
		\end{equation}
		where the disjoint subspaces consist of elements $\tilde \Psi := (\Psi,\vec g_{\vt_c}^{\hspace{0.5mm}\circ},\vec \omega_{\vt_c}^{\hspace{0.5mm}\circ})$ coinciding as set with
		\begin{gather*}
		\hspace{1mm}\cH_{\mathrm{tri}}
		:= \Big\{\tilde \Psi \in \cL^2(\Gamma) :\hspace{1mm}
		v_0=w_0=\eta_0=0,\hspace{1mm} w_s=\eta_s=0,\hspace{1mm}
		v_1=v_2=v_3,
		u_1=u_2=u_3,\\
		\vec g_{\vt_c}^{\hspace{0.5mm}\circ}\cdot\vec E_1 = 0,\hspace{1mm} \vec g_{\vt_c}^{\hspace{0.5mm}\circ}\cdot\vec E_2 = 0, \hspace{1mm} \vec \omega_{\vt_c}^{\hspace{0.5mm}\circ} = 0\Big\},\\
		\hspace{2mm}\cH_{\mathrm{alt}}
		:= \Big\{\tilde \Psi \in \cL^2(\Gamma) :\hspace{1mm}
		v_0=w_0=u_0=0,
		u_s=v_s=0,\hspace{1mm} 
		w_1=w_2=w_3,\hspace{1mm}
		\eta_1=\eta_2=\eta_3,\\
		\vec g_{\vt_c}^{\hspace{0.5mm}\circ} = 0, \hspace{1mm}
		\vec \omega_{\vt_c}^{\hspace{0.5mm}\circ}\cdot\vec E_1 = 0, \hspace{1mm} \vec \omega_{\vt_c}^{\hspace{0.5mm}\circ}\cdot\vec E_2 = 0\Big\},\\
		\cH_\omega
		:= \Big\{\tilde \Psi \in \cL^2(\Gamma) :\hspace{1mm}
		u_0=\eta_0=0,\ b_0w_0=\mathrm{i} a_0v_0,\
		\Psi_3 = \omega \Psi_2
		= \omega^2 \Psi_1,  
		\vec g_{\vt_c}^{\hspace{0.5mm}\circ}\cdot\vec E_3 = 0, \\
		\hspace{30mm}\vec \omega_{\vt_c}^{\hspace{0.5mm}\circ}\cdot\vec E_3 = 0,\hspace{1mm}
		\vec g_{\vt_c}^{\hspace{0.5mm}\circ}\cdot\vec E_2 = \mathrm{i}(\vec g_{\vt_c}^{\hspace{0.5mm}\circ}\cdot\vec E_1), \hspace{1mm}
		\vec \omega_{\vt_c}^{\hspace{0.5mm}\circ}\cdot\vec E_1 = - \mathrm{i}(\vec \omega_{\vt_c}^{\hspace{0.5mm}\circ}\cdot\vec E_2)\Big\} = \cc{\cH_{\cc{\omega}}},
		\end{gather*}
		with $s\in\{1,2,3\}$ labels the legs, $\Psi_s:=(v_s,w_s,u_s,\eta_s)^T$, and $\omega = e^{2\pi\mathrm{i}/3}$.
	\end{thm}
    \begin{proof}[\normalfont \textbf{Proof of Theorem~\ref{thm:reducing_antenna_tower}}] 
    	Decomposition of domain is based on the proof of Theorem 5.1 in \cite{BE21}. In details, the properties of $\Psi_s:=[v_s,w_s,u_s,\eta_s]^T$ in each irreducible subspace is based on condition (i) in the Assumption \ref{symAntenaAssum}. Moreover, properties on the vectors $\vec g_{\vt_c}^{\hspace{0.5mm}\circ}$ and $\vec \omega_{\vt_c}^{\hspace{0.5mm}\circ}$ is based on the edge's force and moments relations \eqref{eq:semiRigid_displacement} and \eqref{eq:semiRigid_rotation} repetitively by setting $e = e_0$ along with application of condition (ii) in the Assumption \ref{symAntenaAssum}.  
    \end{proof}
	\begin{remark}
		\label{3DRemark_1}
		A decomposition $\cH = \prod_{\rho} \cH_\rho$ is
		\textbf{reducing} for an operator $\cA$ if $\cA$ is invariant on each of the subspaces and the operator domain $\Dom(\cA)$ is ``aligned'' with respect to the decomposition, namely 
		\begin{equation*}
			\Dom(\cA) = \prod_{\rho} \big(\cH_\rho\cap\Dom(\cA)\big)
		\end{equation*}
		This means that every aspect of the spectral data of the operator $\cA$ is the union of the spectral data corresponding to a restricted subspace, interested reader may refer to the work \cite{BE21} for detailed background along this line.
	\end{remark}
Abstract result of Theorem \ref{thm:reducing_antenna_tower} maybe applied to explicitly characterize eigenfunctions belonging to each irreducible subspace. In the next example we will restrict such analysis to the case $\cA_{\omega} = \cA|_{\cH_{\omega}}$ and will leave similar calculations for the subspaces $\cH_{\mathrm{tri}}$ and $\cH_{\mathrm{alt}}$ to the interested reader. 

\textbf{Characteristic equation for $\mathbf{\cA_{\omega}}$ operator}. Restricting to the $\cH_\omega$ subspace, imposing free vertex condition at $\vt_0$ and fixed vertex conditions at $\vt_1$, $\vt_2$, and $\vt_3$, then general solution set is of the form 
\begin{gather*}
	\hspace{1.25mm}v(x) = A_v(\sinh(\mu x) - \sin(\mu x)) + B_v (\cosh(\mu x) - \cos(\mu x)), \\
	iw(x) = A_w(\sinh(\mu x) - \sin(\mu x)) + B_w(\cosh(\mu x) - \cos(\mu x)),\\
	\hspace{1.5mm}u(x) = A_u\sin(\beta x),\\
	i\eta(x) = A_\eta \sin(\beta x),\\
	v_0(x) = A_0(\sinh(\mu x) + \sin(\mu x))
	+ B_0(\cosh(\mu x) + \cos(\mu x)),
\end{gather*}
where $\mu = (\lambda/a)^{1/4}$, and $\beta = (\lambda/d)^{1/2}$. By the assumption that discontinuity of fields are only admissible at $\vt_c$,  derivation of vertex conditions are discussed in the Appendix. Constraining to the case in which $\kappa_{g_1}^{-1}, \kappa_{\omega_1}^{-1} \rightarrow 0$, and assuming unit material parameters $a=a_0=d=1$, then application of set linearly independent conditions at the central vertex $\vt_c$ summarized in remark \ref{linearDependt},  eigenvalue problem reduces to finding the ($n$-dependent) coefficient vector $[A_v, B_v, A_w, B_w, A_u, A_\eta, A_0, B_0, (\vec g_{\vt_c}^{\hspace{0.5mm}\circ}\cdot \vec E_1), (\vec \omega_{\vt_c}^{\hspace{0.5mm}\circ} \cdot \vec E_2)]^\text{T}$ in the kernel of the $10\times10$ matrix $\mathbb{M}(\lambda;\kappa_{g_0}^{-1}, \kappa_{\omega_0}^{-1})$ decomposed as
\begin{equation}
	\label{eq:MRJ}
	\mathbb{M}(\lambda;\kappa_{g_0}^{-1}, \kappa_{\omega_0}^{-1}) = \mathbb{M}_r(\lambda) + \mathbb{M}_{g_0} (\lambda;\kappa_{g_0}^{-1}) + \mathbb{M}_{\omega_0} (\lambda;\kappa_{\omega_0}^{-1})
\end{equation}
Above, modification towards admissible discontinuity of displacement and rotation fields at $\vt_c$ are given by matrices $\mathbb{M}_{g_0}$ and $\mathbb{M}_{\omega_0}$, respectively defined as
\begin{subequations}
	\label{eq:Jump0}
	\begin{equation}
	\mathbb{M}_{g_0}(\lambda;\kappa_{g_0}^{-1})_{p,q} = -\kappa_{g_0}^{-1}\mu^3(C^-_{\mu}\delta_{q,7} + S^+_{\mu} \delta_{q,8})\delta_{7,p} 
	\end{equation}
	\begin{equation}
	\mathbb{M}_{\omega_0}(\lambda;\kappa_{\omega_0}^{-1})_{p,q} = +\kappa_{\omega_0}^{-1}\mu^2(S^-_{\mu}\delta_{q,7} + C^-_{\mu} \delta_{q,8})\delta_{8,p} 
	\end{equation}
\end{subequations}
while matrix $\mathbb{M}_r(\lambda)$ governs the problem with rigid joint model $\vt_c$ with entries\footnote{See section 5.3.3 in \cite{BE21} for an alternative derivation of characteristic equation corresponding rigid-joint model.}
\begin{myequation1}
	\begin{pmatrix}
			\mu C^-_{\mu} & \mu S^+_{\mu} & 0 & 0 & 0 & 0 & 0 & 0 & \hspace{3mm}0 & 1\\
			S^-_{\mu}S_\alpha & C^-_{\mu}S_\alpha & 0 & 0 & -S_{\mu^2}C_\alpha & 0 & 0 & 0 & \hspace{3mm}1 & 0\\
			S^-_{\mu}C_\alpha & C^-_{\mu}C_\alpha & 0 & 0 & +S_{\mu^2}S_\alpha & 0 & 0 & 0 & \hspace{3mm}0 & 0\\
			0 & 0 & S^-_{\mu} & C^-_{\mu} & 0 & 0 & 0 & 0 & \hspace{3mm}1 & 0\\
			0 & 0 & \mu C^-_{\mu}S_\alpha & \mu S^+_{\mu}S_\alpha & 0 & -S_{\mu^2}C_\alpha & 0 & 0 & \hspace{3mm}0 & 1\\
			0 & 0 & \mu C^-_{\mu}C_\alpha & \mu S^+_{\mu}C_\alpha & 0 & +S_{\mu^2}S_\alpha & 0 & 0 & \hspace{3mm}0 & 0\\
			0 & 0 & 0 & 0 & 0 & 0 & S^+_{\mu}& C^+_{\mu} & -1 & 0\\
			0 & 0 & 0 & 0 & 0 & 0 & \mu C^+_{\mu}& \mu S^-_{\mu} & \hspace{3mm}0 & 1\\
			\frac{3}{2}\mu C^+_{\mu}S_\alpha & \frac{3}{2}\mu S^-_{\mu}S_\alpha & \frac{3}{2}\mu C^+_{\mu} & \frac{3}{2}\mu S^-_{\mu}& \frac{3}{2}C_{\mu^2}C_\alpha & 0 & -\mu C^-_{\mu} & -\mu S^+_{\mu} & -\mu^2 m_\vt & 0\\
			\frac{3}{2}S^+_{\mu} & \frac{3}{2} C^+_{\mu} & \frac{3}{2}S^+_{\mu}S_\alpha & \frac{3}{2} C^+_{\mu}S_\alpha & 0 & -\frac{3}{2}C_{\mu^2}C_\alpha &  S^-_{\mu} & C^-_{\mu} & \hspace{3mm}0 & -\mu^2 m_\vt
	\end{pmatrix}
\end{myequation1}
\section{Local Spectral Basis and Graph Geometry}
\label{sec:CompactGraph2D}
\subsection{Preliminary}
\begin{figure}[ht]
	\centering
	\includegraphics[width=0.45\textwidth]{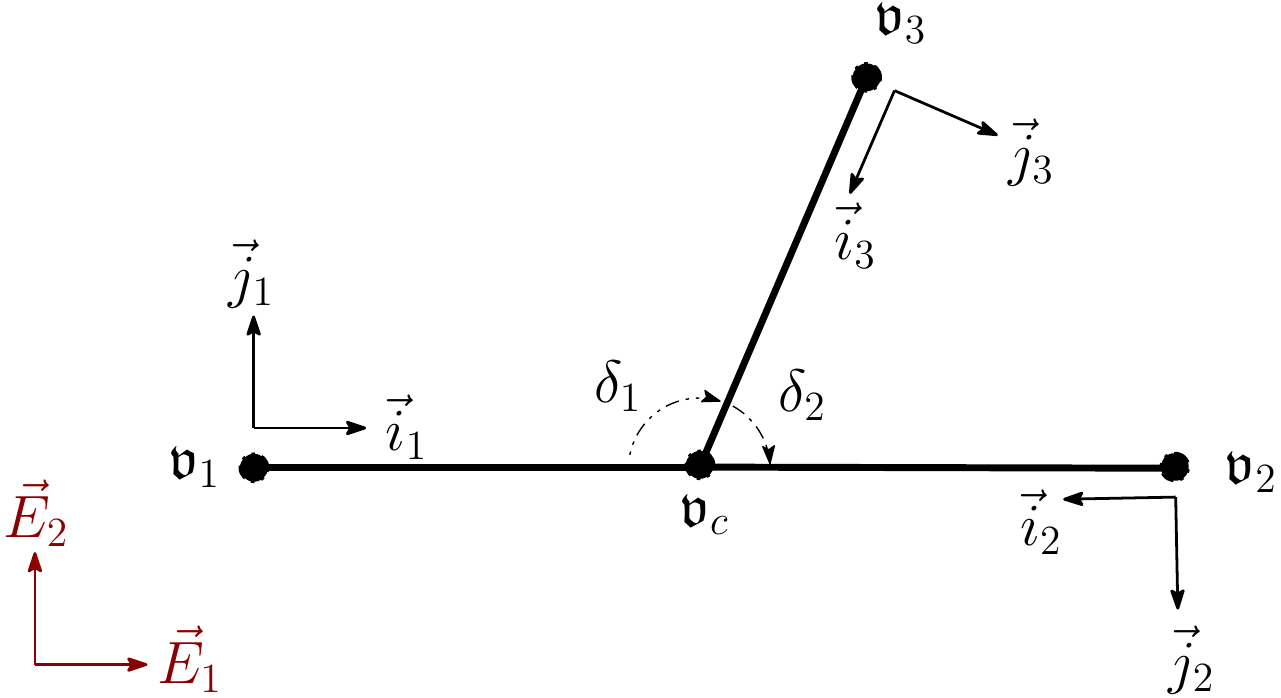}
	\caption{Geometry of 3-star planar graph in its equilibrium
		sate. Variation of angles $\delta_1$ and $\delta_2$ span all possible geometries corresponding to this class of graphs.} 
	\label{fig:2DGraph}
\end{figure}

The goal of this Section is to develop characteristics equation corresponding to an eigenvalue problem on 3-star planar graph $\Gamma$, see Figure \ref{fig:2DGraph}. Unlike the former examples, here this task will be done by introducing geometric-free local spectral basis applicable to each edge of graph, while geometry of $\Gamma$ will be later encoded in the coefficient set of constructed fields. This method is general and clearly can be applied to derive characteristic equation derived in the former examples. But we keep this ordering of exhibition to emphasize the benefits of each method. In contrast with the commonly applied (fixed) local basis in numerical methods, e.g. finite element method\cite{AB17}, here each basis solves an eigenvalue problem on a fixed segment $[0,1]$ with appropriate boundary conditions. This can be considered as an extension of the framework applied in analysis of periodic graphs in \cite{KP07}, by considering coupled fields and enriching solution's domain in which discontinuity of fields are admissible. For the sake of clearness, discussion will be limited to a special case in which the boundary vertices $\vt_1, \vt_2$, and $\vt_3$ are fixed. More general cases will be a straightforward extension and will be presented on analysis of periodic graphs in the forthcoming manuscript \cite{ES21}. 

Consider a \emph{planar} frame depicted in Figure
\ref{fig:2DGraph} consisting of three beams $e_1, e_2, e_3$ meeting at the free central semi-rigid joint $\vt_c$. The beams are oriented from the fixed ends to the joint; the local basis of each beam is shown in the figure. Introducing $\Psi_e(x) := [v_e(x), \eta_e(x)]^{\text{T}}$, we are interested on eigenvalue problem \eqref{eq:outPlane} on this graph. The vertex conditions in Corollary \ref{decouplingPlane} implies that at each vertex $\vt$ and for each $e \sim \vt$, subjected to displacement-force condition
\begin{equation}
	\label{eq:vertexCondPlanarPrimary_V_1}
	v_e(\vt) + s_e^\vt a_e \kappa_{g_v}^{-1} v_e''' (\vt)= v_\vt^\circ
\end{equation}
and rotation-moment conditions
\begin{equation}
	\label{eq:vertexCondPlanarPrimary_V_2}
	\eta_e(\vt) + s_e^\vt d_e\kappa_{\omega_\eta}^{-1} \eta_e'(\vt) =  \vec \omega_{\vt}^{\hspace{0.5mm}\circ} \cdot \vec i_e, \qquad
	- v_e'(\vt) - s_e^\vt a_e\kappa_{\omega_v}^{-1}  v_e''(\vt) =  \vec \omega_{\vt}^{\hspace{0.5mm}\circ} \cdot \vec j_e
\end{equation}
Moreover, net-force and moments at $\vt$ implies that 
\begin{equation}
\label{eq:forceMom}
\begin{split}
-\displaystyle\sum_{e \sim \vt} s_e^\vt a_e v_e'''(\vt)  = \lambda \mathfrak{m}_\vt v_{\vt}^\circ, \qquad 
\displaystyle\sum_{e \sim \vt} s_e^\vt \big(d_e \eta_e'(\vt) \vec i_e - a_e v_e''(\vt) \vec j_e\big) = \lambda \mathfrak{m}_\vt 
\vec \omega_{\vt}^{\hspace{0.5mm}\circ}
\end{split}
\end{equation}
In this Section we will assume that boundary vertices $\vt_1, \vt_2$ and $\vt_3$ are rigid, i.e. the vectors $v_s(0) = v_{\vt_s}^\circ$ and $\vec \omega_s(0) = \vec \omega_{\vt_s}^{\hspace{0.5mm}\circ}$ for $s=1,2,3$. Semi-rigid vertex model is assumes at the central vertex $\vt_c$, on which conditions \eqref{eq:vertexCondPlanarPrimary_V_1} and \eqref{eq:vertexCondPlanarPrimary_V_2} for unit length edges, $\ell_e = 1$, reduce to
\begin{gather}
v_1(1) + a \kappa_{g_v}^{-1} v_1'''(1) = v_2(1) + a \kappa_{g_v}^{-1} v_2'''(1) = v_3(1) + a \kappa_{g_v}^{-1} v_3'''(1) = v_{\vt_c}^\circ,
\end{gather}
and
\begin{subequations}
\begin{gather}
\eta_1(1) + d \kappa_{\omega_\eta} ^{-1} \eta_1'(1) = \vec \omega_{\vt_c}^{\hspace{0.5mm}\circ} \cdot \vec i_1, \qquad 
-v_1'(1) - a \kappa_{\omega_v} ^{-1} v_1''(1)  = \vec \omega_{\vt_c}^{\hspace{0.5mm}\circ} \cdot \vec j_1
\\
\eta_2(1) + d \kappa_{\omega_\eta} ^{-1} \eta_2'(1) = \vec \omega_{\vt_c}^{\hspace{0.5mm}\circ}  \cdot \vec i_2, \qquad 
-v_2'(1) - a \kappa_{\omega_v} ^{-1} v_2''(1)  = \vec \omega_{\vt_c}^{\hspace{0.5mm}\circ} \cdot \vec j_2 
\\
\eta_3(1) + d \kappa_{\omega_\eta} ^{-1} \eta_3'(1) = \vec \omega_{\vt_c}^{\hspace{0.5mm}\circ}  \cdot \vec i_3, \qquad 
-v_3'(1) - a \kappa_{\omega_v} ^{-1} v_3''(1)  = \vec \omega_{\vt_c}^{\hspace{0.5mm}\circ}  \cdot \vec j_3
\end{gather}
\end{subequations}
Expansion of conditions \eqref{eq:forceMom} edgewise, then at the central vertex 
\begin{subequations}
\begin{gather}
-a v_1'''(1) - a v_2'''(1) - a v_3'''(1)= \lambda \mathfrak{m}_{\vt_c}  v_{\vt_c}^\circ,\\
d\eta_1'(1) \vec i_1 + d \eta_2'(1) \vec i_2 + d \eta_3'(1) \vec i_3 -a v_1''(1)\vec j_1 -a v_2''(\ell_2)\vec j_2 -a v_3''(1)\vec j_3 = \lambda \mathfrak{m}_{\vt_c} \vec \omega_{\vt_c}^{\hspace{0.5mm}\circ}
\end{gather}
Next we will construct geometric-free local spectral basis (eigenvalue dependent) for each field $\eta(x)$ and $v(x)$ separately.
\end{subequations}  

\subsection{Local Representation}
\subsubsection{Spectral Basis For Out of Plane Displacement} 
Let denote by $\Sigma_v^\text{D}$ to be the spectrum of an operator $\cA_v v(x) = av''''(x) $ on an interval $[0,1]$ with fixed and Robin boundary conditions respectively at $x = 0$ and $x=1$, i.e. $v(x)$ satisfying
\begin{equation}
\label{eq:DRB_v}
\begin{split}
v(0) = 0, \qquad  v'(0) = 0, \qquad  
v(1) + a\kappa_{g_v}^{-1}v'''(1) = 0, \qquad  v'(1) + a\kappa_{\omega_v}^{-1}v''(1) = 0.
\end{split}
\end{equation}
Imposing boundary conditions \eqref{eq:DRB_v}, then set $\Sigma_\eta^\text{D}$ is equivalent to 
\begin{equation}
\label{eq:sigmDeta_v1}
\Sigma_v^\text{D} := \big\{\lambda \in \bR ~:~  \cD_v(\lambda;\kappa_{g_v}^{-1},\kappa_{\omega_v}^{-1},a) = 0 \big\}
\end{equation}
where for $\mu_a := (\lambda/a)^{1/4}$, function $\cD_v(\lambda;\kappa_{g_v}^{-1},\kappa_{\omega_v}^{-1},a)$ is of a form
\begin{equation}
\cD_v = 
(S^-_\mu-a \kappa_{g_v}^{-1}\mu^3C^+_\mu)(S^+_\mu+a \kappa_{\omega_v}^{-1}\mu C^+_\mu)-
(C^-_\mu-a \kappa_{g_v}^{-1}\mu^3 S^-_\mu)(C^-_\mu+a \kappa_{\omega_v}^{-1}\mu S^+_\mu)
\end{equation}
For the class of fixed boundary conditions at both boundary vertices, $\cD_v = 2(\cosh(\mu_a)\cos(\mu_a)+1)$ which is well known characteristic equation\footnote{For example see \href{https://en.wikipedia.org/wiki/Euler-Bernoulli_beam_theory}{wikipedia page}.} for the so-called clamped cantilevered beam with unit length, see \eqref{eq:1DSeq} for the case of fixed/free boundaries counterpart. If $\lambda \notin \Sigma_v^\text{D} \cup \{0\}$, there exist four linearly independent solutions $\phi_1, \phi_2,  \phi_3$ and $ \phi_4$ on $[0,1]$ of
\begin{equation}
\label{eq:Eig_V}
\cA_v \phi_k(x) = a\phi_k''''(x)  = \lambda\phi_k(x)
\end{equation}
such that at the boundary points satisfying
\begin{equation}
\label{eq:PhiEqs}
\begin{split}
\phi_1(0) &= 1, \qquad \phi_1'(0) = 0, \qquad \phi_1(1) - a\kappa_{g_v}^{-1}\phi_1'''(1) = 0, \qquad \phi_1'(1) +a\kappa_{\omega_v}^{-1}\phi_1''(1) = 0, \\
\phi_2(0) &= 0, \qquad \phi_2'(0) = 1, \qquad \phi_2(1) - a\kappa_{g_v}^{-1}\phi_2'''(1) = 0, \qquad  \phi_2'(1) +a\kappa_{\omega_v}^{-1}\phi_2''(1) = 0,\\
\phi_3(0) &= 0, \qquad \phi_3'(0) = 0, \qquad \phi_3(1) - a\kappa_{g_v}^{-1}\phi_3'''(1) = 1, \qquad  \phi_3'(1) +a\kappa_{\omega_v}^{-1}\phi_3''(1) = 0,\\
\phi_4(0) &= 0, \qquad \phi_4'(0) = 0, \qquad \phi_4(1) - a\kappa_{g_v}^{-1}\phi_4'''(1) = 0, \qquad  \phi_4'(1) +a\kappa_{\omega_v}^{-1}\phi_4''(1) = 1.
\end{split}
\end{equation}
Thereby, for $\lambda \not \in \Sigma_v^\text{D} \cup \{0\}$ and for each $k=1,\ldots,4$, general solution $\phi_k(x)$ is of a form
\begin{equation}
\label{eq:solPhiG1}
\phi_k(x) = A_k \sinh(\mu_a x) + B_k \sin(\mu_a x) + C_k \cosh(\mu_a x) + D_k \cos(\mu_a x)
\end{equation}
Imposing vertex conditions \eqref{eq:PhiEqs} determines the local basis $\phi_k(x)$, e.g.  
\begin{equation}
\label{eq:DvSol}
\begin{split}
\phi_3(x) &= \frac{1}{\cD_v(\lambda;\kappa_{g_v}^{-1},\kappa_{\omega_v}^{-1},a)} \big((S^+_{\mu_a}+a \kappa_{\omega_v}^{-1}{\mu_a} C^+_{\mu_a})S^-_{{\mu_a} x}-(C^-_{\mu_a}+a \kappa_{\omega_v}^{-1}{\mu_a} S^+_{\mu_a}) C^-_{{\mu_a} x} \big),\\
\phi_4(x) &= \frac{-\mu_a^{-1}}{\cD_v(\lambda;\kappa_{g_v}^{-1},\kappa_{\omega_v}^{-1},a)} \big((C^-_{\mu_a}-a \kappa_{g_v}^{-1}{\mu_a^3} S^-_{\mu_a})S^-_{{\mu_a} x}-(S^-_{\mu_a}-a \kappa_{g_v}^{-1}{\mu_a^3}C^+_{\mu_a}) C^-_{{\mu_a} x} \big).
\end{split}
\end{equation}
where the notations above are introduced in \eqref{eq:notationMatrix}. This set of basis will be later applied on writing the general solution for fields $v_e(x)$. We stress that no-information regarding the geometry exists in the derived basis set as they are purely local.  
\subsubsection{Spectral Basis For Angular Displacement}
\begin{figure}[ht]
	\centering
	\includegraphics[width=0.95\textwidth]{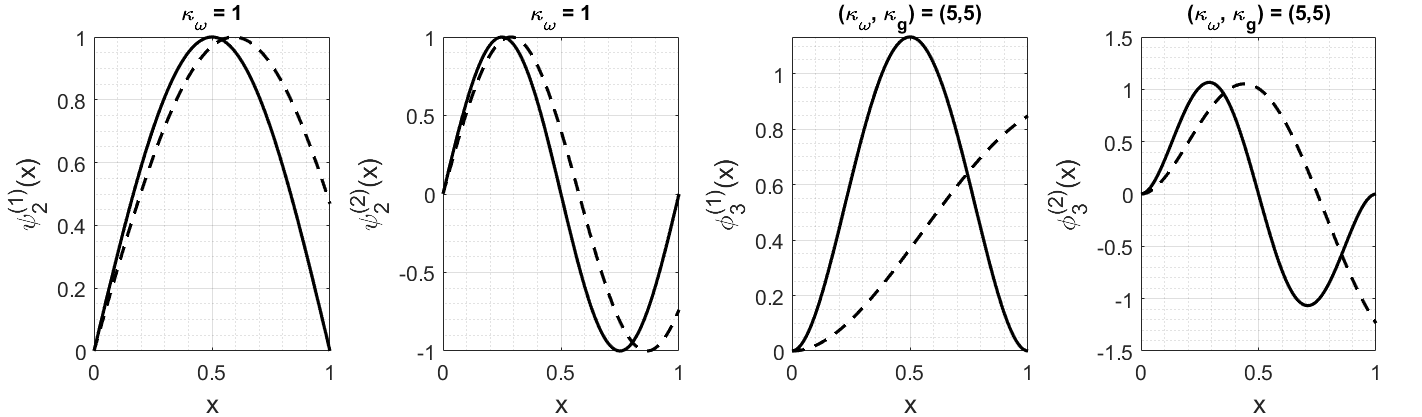}
	\caption{Plots of first two eigenfunctions corresponding to sets $\Sigma_\eta^\text{D}$ and $\Sigma_v^\text{D}$. The solid lines corresponds to the rigid-vertex case in which fixed boundary conditions are applied on both end of the domain $[0,1]$.}
	\label{fig:LocalBasisPlot}
\end{figure}

Similar concept can be applied to derive local basis for $\eta(x)$ field. This will be similar to the derivation in \cite{KP07} with the difference of Robin type boundary condition on one end. In fact, let denote by $\Sigma_\eta^\text{D}$ to be the solution to an eigenvalue problem $\cA_\eta \eta(x) = -d\eta''(x) = \lambda \eta(x)$ on interval $[0,1]$ with boundary conditions 
\begin{equation}
\label{eq:DRB_E}
\begin{split}
\eta(0) = 0, \qquad \eta(1) + d\kappa_{\omega_\eta}^{-1}\eta'(1) = 0
\end{split}
\end{equation}
Imposing the fixed/Robin boundary conditions \eqref{eq:DRB_E}, this set $\Sigma_\eta^\text{D}$ is equivalent to
\begin{equation}
\label{eq:sigmDeta}
\Sigma_\eta^\text{D} := \big\{\lambda \in \bR ~:~ \cD_\eta(\lambda;\kappa_{\omega_\eta}^{-1},d) = 0 \big\}
\end{equation}
where for $\beta_d := (\lambda/d)^{1/2}$, function  $\cD_\eta(\lambda;\kappa_{\omega_\eta}^{-1},d)$ is of a form 
\begin{equation}
\cD_\eta := \sin(\beta_d) + d\kappa_{\omega_\eta}^{-1}\beta_d\cos(\beta_d)
\end{equation}
If $\lambda \notin \Sigma_\eta^\text{D} \cup \{0\}$, there exist two linearly independent solutions $\psi_1(x)$ and $ \psi_2(x)$, depending on $\lambda$, on $[0,1]$ of eigenvalue problem
\begin{equation}
\label{eq:Eig_E}
\cA_\eta \psi(x) = -d\psi''(x)  = \lambda\psi(x)
\end{equation}
such that at boundary points of the interval satisfy
\begin{equation}
\label{eq:PsiEqs}
\begin{split}
\psi_1(0) &= 1, \qquad \psi_1(1) + d \kappa_{\omega_\eta}^{-1}\psi_1'(1) = 0, \\
\psi_2(0) &= 0, \qquad \psi_2(1) + d \kappa_{\omega_\eta}^{-1}\psi_2'(1) = 1.
\end{split}
\end{equation}
For the case that $\lambda \not \in \Sigma_\eta^\text{D} \cup \{0\}$, general solution of eigenvalue problem \eqref{eq:Eig_E} can be written as
\begin{equation}
\label{eq:solPhi}
\psi_k(x) = A_k \sin(\beta_d x) + B_k \cos(\beta_d x)
\end{equation}
Imposing vertex conditions \eqref{eq:PsiEqs}, then for such $\lambda$'s, basis functions can be explicitly determined, e.g. 
\begin{equation}
\label{eq:psi2}
\psi_2(x) =  \frac{1}{\cD_\eta(\lambda;\kappa_{\omega_\eta}^{-1},d)} \sin(\beta_d x) 
\end{equation}
Figure \ref{fig:LocalBasisPlot} plots functions $\psi_2(x)$, $\phi_3(x)$, and $\phi_4(x)$ corresponding to the first two elements of sets $\Sigma_\eta^\text{D}$ and $\Sigma_v^\text{D}$ respectively. One may observes that if $\lambda \in \Sigma_v^\text{D}\cup \{0\}$, then solution $v_e(x)$ for all $e$ will be decoupled and can be determined explicitly, e.g. see \eqref{eq:DvSol}. Similar concept holds for $\eta_e(x)$.
\subsection{Geometry via Global Representation}
In the following we will construct solution $\Psi(x)$ to the coupled eigenvalue problem on the 3-star graph using the local basis developed above. Key idea on bringing geometry of the graph into play is by writing the solution for fields as a linear combination of local basis in such a way that they automatically satisfy vertex conditions \eqref{eq:vertexCondPlanarPrimary_V_1} and \eqref{eq:vertexCondPlanarPrimary_V_2}. We will assume that the ($n$-dependent) basis functions $\phi_k(x)$ and $\psi_k(x)$ are lifted to each of the edges in graph $\Gamma$, using the described before identifications of these edges with the segment $[0, 1]$. Abusing notations, we will use the same names $\phi_k$ and $\psi_k$ for the lifted functions. Denoting by 
\begin{equation}
\label{eq:sigMD}
\Sigma^\text{D} := \Sigma_v^\text{D} \cup \Sigma_\eta^\text{D} \cup \{0\},
\end{equation}
then for $\lambda \not \in \Sigma^\text{D}$ we can use \eqref{eq:Eig_V} and \eqref{eq:Eig_E} to represent any solution $\Psi(x)$ of eigenvalue problem \eqref{eq:outPlane} on each edge $e$. In detail, solution $v_s(x)$ for edge $e_s$ and $s=1,2,3$ will be represented as
\begin{equation}
\label{eq:vR}
\begin{split}
v_1(x) = v_{\vt_1}^\circ \phi_1(x) - (\vec \omega_{\vt_1}^{\hspace{0.5mm}\circ} \cdot \vec j_1)\phi_2(x) + 
v_{\vt_c}^\circ \phi_3(x) - (\vec \omega_{\vt_c}^{\hspace{0.5mm}\circ} \cdot \vec j_1)\phi_4(x), \\
v_2(x) = v_{\vt_2}^\circ \phi_1(x) - (\vec \omega_{\vt_2}^{\hspace{0.5mm}\circ} \cdot \vec j_2)\phi_2(x) + 
v_{\vt_c}^\circ \phi_3(x) - (\vec \omega_{\vt_c}^{\hspace{0.5mm}\circ} \cdot \vec j_2)\phi_4(x),
\\
v_3(x) = v_{\vt_3}^\circ \phi_1(x) - (\vec \omega_{\vt_3}^{\hspace{0.5mm}\circ} \cdot \vec j_3)\phi_2(x) + 
v_{\vt_c}^\circ \phi_3(x) - (\vec \omega_{\vt_c}^{\hspace{0.5mm}\circ} \cdot \vec j_3)\phi_4(x).
\end{split}
\end{equation}
while solutions $\eta_s(x)$ are of the form
\begin{equation}
\label{eq:eR}
\begin{split}
\eta_1(x) = (\vec \omega_{\vt_1}^{\hspace{0.5mm}\circ}  \cdot \vec i_1) \psi_1(x) +
(\vec \omega_{\vt_c}^{\hspace{0.5mm}\circ} \cdot \vec i_1) \psi_2(x), \\
\eta_2(x) = (\vec \omega_{\vt_2}^{\hspace{0.5mm}\circ} \cdot \vec i_2) \psi_1(x) + 
(\vec \omega_{\vt_c}^{\hspace{0.5mm}\circ} \cdot \vec i_2) \psi_2(x), \\
\eta_3(x) = (\vec \omega_{\vt_3}^{\hspace{0.5mm}\circ} \cdot \vec i_3) \psi_1(x) + 
(\vec \omega_{\vt_c}^{\hspace{0.5mm}\circ} \cdot \vec i_3) \psi_2(x).
\end{split}
\end{equation}
Interested reader may try to check vertex conditions \eqref{eq:vertexCondPlanarPrimary_V_1} and \eqref{eq:vertexCondPlanarPrimary_V_2} by replacing the representations \eqref{eq:vR} and \eqref{eq:eR} along with applying boundary properties of local spectral basis stated in \eqref{eq:PhiEqs} and \eqref{eq:PsiEqs}. We stress that that for a fixed graph's geometry, the unknown coefficients consist of scalar quantity $v_{\vt_c}^\circ$ and vector $\vec \omega_{\vt_c}^{\hspace{0.5mm}\circ}$ of size two. The latter has representation in the global coordinate as
\begin{equation}
	\vec \omega_{\vt_c}^{\hspace{0.5mm}\circ} = 
	(\vec \omega_{\vt_c}^{\hspace{0.5mm}\circ} \cdot \vec E_1) \vec E_1 + (\vec \omega_{\vt_c}^{\hspace{0.5mm}\circ} \cdot \vec E_2) \vec E_2 
\end{equation}
Moreover, for $\ell=1,2$, let introduce two vectors $\mathbf{I}_{\text{E}_\ell}$ and $\mathbf{J}_{\text{E}_\ell}$ which encode geometric information of the graph through 
\begin{equation}
\label{eq:IJs}
\mathbf{I}_{\text{E}_\ell} := [(\vec E_\ell \cdot \vec i_1),(\vec E_\ell \cdot \vec i_2),(\vec E_\ell \cdot \vec i_3)]^{\text{T}}, \qquad
\mathbf{J}_{\text{E}_\ell} := [\vec E_\ell \cdot \vec j_1),(\vec E_\ell \cdot \vec j_2),(\vec E_\ell \cdot \vec j_3)]^{\text{T}}
\end{equation}
Application of presentations \eqref{eq:vR} and \eqref{eq:eR} in the net force and moment relations \eqref{eq:forceMom}, reduces the problem to the following set of linearly independent conditions at the central vertex $\vt_c$ of a form
\begin{equation*}
	\hspace{1mm}-3a \phi_3'''(1) v_{\vt_c}^\circ + 
	 a(\mathbf{1}^{\text{T}}\mathbf{J}_{\text{E}_1})\phi_4'''(1)(\vec \omega_{\vt_c}^{\hspace{0.5mm}\circ} \cdot \vec E_1) + a(\mathbf{1}^{\text{T}}\mathbf{J}_{\text{E}_2})\phi_4'''(1)(\vec \omega_{\vt_c}^{\hspace{0.5mm}\circ} \cdot \vec E_2) = \lambda \mathfrak{m}_{\vt_c} v_{\vt_c}^\circ,
\end{equation*}
\begin{equation*}
\begin{split}
-a(\mathbf{1}^{\text{T}}\mathbf{J}_{\text{E}_1})\phi_3''(1)v_{\vt_c}^\circ + 
a(\mathbf{J}_{\text{E}_1}^{\text{T}} \mathbf{J}_{\text{E}_1})\phi_4''(1)(\vec \omega_{\vt_c}^{\hspace{0.5mm}\circ} \cdot \vec E_1) &+
a(\mathbf{J}_{\text{E}_1}^{\text{T}} \mathbf{J}_{\text{E}_2})\phi_4''(1)(\vec \omega_{\vt_c}^{\hspace{0.5mm}\circ} \cdot \vec E_2) +\\
d(\mathbf{I}_{\text{E}_1}^{\text{T}} \mathbf{I}_{\text{E}_1})\psi_2'(1)(\vec \omega_{\vt_c}^{\hspace{0.5mm}\circ} \cdot \vec E_1) &+ d(\mathbf{I}_{\text{E}_1}^{\text{T}} \mathbf{I}_{\text{E}_2})\psi_2'(1)(\vec \omega_{\vt_c}^{\hspace{0.5mm}\circ} \cdot \vec E_2) = \lambda \mathfrak{m}_{\vt_c} (\vec \omega_{\vt_c}^{\hspace{0.5mm}\circ} \cdot \vec E_1),
\end{split}
\end{equation*}
and
\begin{equation*}
\begin{split}
-a(\mathbf{1}^{\text{T}}\mathbb{J}_{\text{E}_2})\phi_3''(1)v_{\vt_c}^\circ + 
a(\mathbf{J}_{\text{E}_2}^{\text{T}} \mathbf{J}_{\text{E}_1})\phi_4''(1)(\vec \omega_{\vt_c}^{\hspace{0.5mm}\circ} \cdot \vec E_1) &+
a(\mathbf{J}_{\text{E}_2}^{\text{T}} \mathbf{J}_{\text{E}_2})\phi_4''(1)(\vec \omega_{\vt_c}^{\hspace{0.5mm}\circ} \cdot \vec E_2)+ \\
d(\mathbf{I}_{\text{E}_2}^{\text{T}} \mathbf{I}_{\text{E}_1})\psi_2'(1)(\vec \omega_{\vt_c}^{\hspace{0.5mm}\circ} \cdot \vec E_1) &+ d(\mathbf{I}_{\text{E}_2}^{\text{T}} \mathbf{I}_{\text{E}_2})\psi_2'(1)(\vec \omega_{\vt_c}^{\hspace{0.5mm}\circ} \cdot \vec E_2) = \lambda \mathfrak{m}_{\vt_c} (\vec \omega_{\vt_c}^{\hspace{0.5mm}\circ} \cdot \vec E_2).
\end{split}
\end{equation*}
These then reduce the eigenvalue problem to finding the ($n$-dependent) coefficient vector $[v_{\vt_c}^\circ, (\vec \omega_\vt^{\hspace{0.5mm}\circ} \cdot \vec E_1),  (\vec \omega_\vt^{\hspace{0.5mm}\circ} \cdot \vec E_2)]^{\text{T}}$ in the kernel of the $3\times3$ matrix $\mathbb{M}(\lambda)$ defined as
\begin{equation}
\label{eq:M2D}
	\mathbb{M}(\lambda;\kappa_{g_v}^{-1},\kappa_{\omega_\vt}^{-1},\kappa_{\omega_\eta}^{-1},\mathfrak{m}_\vt) := a \mathbb{M}_v(\lambda;\kappa_{g_v}^{-1},\kappa_{\omega_\vt}^{-1}) + d \mathbb{M}_\eta(\lambda;\kappa_{\omega_\eta}^{-1}) - \lambda \mathfrak{m}_\vt \mathbb{I}
\end{equation}
with constituent matrices 
\begin{equation*}
\mathbb{M}_v := 
\begin{pmatrix}
-\phi_3'''(1)\mathbf{1}^{\text{T}}\mathbf{1} &\phi_4'''(1)\mathbf{1}^{\text{T}}\mathbf{J}_{\text{E}_1} &     \phi_4'''(1)\mathbf{1}^{\text{T}}\mathbf{J}_{\text{E}_2}\\
\hspace{3mm}-\phi_3''(1) \mathbf{1}^{\text{T}}\mathbf{J}_{\text{E}_1}& \phi_4''(1)\mathbf{J}_{\text{E}_1}^{\text{T}} \mathbf{J}_{\text{E}_1} & 
\phi_4''(1) \mathbf{J}_{\text{E}_1}^{\text{T}} \mathbf{J}_{\text{E}_2} \\
\hspace{3mm}-\phi_3''(1)\mathbf{1}^{\text{T}}\mathbf{J}_{\text{E}_2} & \phi_4''(1)\mathbf{J}_{\text{E}_2}^{\text{T}} \mathbf{J}_{\text{E}_1} & 
\phi_4''(1)\mathbf{J}_{\text{E}_2}^{\text{T}} \mathbf{J}_{\text{E}_2}
\end{pmatrix}, \quad 
\mathbb{M}_\eta := 
\begin{pmatrix}
0 & 0 & 0 \\
0 & \psi_2'(1)\mathbf{I}_{\text{E}_1}^{\text{T}} \mathbf{I}_{\text{E}_1} & 
\psi_2'(1)\mathbf{I}_{\text{E}_1}^{\text{T}} \mathbf{I}_{\text{E}_2} \\
0 & \psi_2'(1)\mathbf{I}_{\text{E}_2}^{\text{T}} \mathbf{I}_{\text{E}_1} & 
\psi_2'(1)\mathbf{I}_{\text{E}_2}^{\text{T}} \mathbf{I}_{\text{E}_2}
\end{pmatrix}
\end{equation*}
where $\mathbf{1}$ is a vector of size 3 with unit entries. The graph's geometric information in \eqref{eq:M2D} can be further factorized from the local basis set. We will summarize the derivations above in the following result.      
\begin{prop}
	\label{Prop2D_1}
	Eigenvalues $\lambda \not \in \Sigma^\text{D}$ corresponding to Hamiltonian \eqref{eq:outPlane} on 3-star planar graph with (rigid) fixed boundary conditions and semi-rigid central vertex is the solution of characteristic equation $\det(\mathbb{M}(\lambda)) = 0$ where 
	\begin{equation}
	\label{eq:M2DExp}
	\mathbb{M}(\lambda) = -a\big(\phi_3'''(1)\mathbb{G}_\mathbf{0}+\phi_3''(1)\mathbb{G}_\mathbf{1} -\phi_4'''(1)\mathbb{G}_\mathbf{1}^{\text{T}} - \phi_4''(1)\mathbb{G}_\mathbf{J} \big) + d \psi_2'(1) \mathbb{G}_\mathbf{I} - \lambda \mathfrak{m}_\vt \mathbb{I}
	\end{equation}
	with $\mathbb{G}_\mathbf{0}$ is a $3 \times 3$ matrix will all its entries zero except $(\mathbb{G}_0)_{11} = \mathbf{1}^{\text{T}}\mathbf{1}$. Moreover, $\mathbb{M}$ is constructed out of local spectral basis defined in \eqref{eq:Eig_V}, \eqref{eq:Eig_E}, and purely geometric (global) basis 
	\begin{equation*}
	\mathbb{G}_\mathbf{1} :=  
	\begin{pmatrix}
		0 & 0 & 0 \\ 
		\mathbf{1}^{\text{T}}\mathbf{J}_{\text{E}_1} & 0 & 0 \\ 
		\mathbf{1}^{\text{T}}\mathbf{J}_{\text{E}_2} & 0 & 0
	\end{pmatrix}, \quad 
	\mathbb{G}_\mathbf{J} :=  
	\begin{pmatrix}
	0 & 0 & 0 \\
	0 & \mathbf{J}_{\text{E}_1}^{\text{T}} \mathbf{J}_{\text{E}_1} & \mathbf{J}_{\text{E}_1}^{\text{T}} \mathbf{J}_{\text{E}_2} \\ 
	0 & \mathbf{J}_{\text{E}_2}^{\text{T}} \mathbf{J}_{\text{E}_1} & \mathbf{J}_{\text{E}_2}^{\text{T}} \mathbf{J}_{\text{E}_2}
	\end{pmatrix}, \quad 
	\mathbb{G}_\mathbf{I} :=  
	\begin{pmatrix}
		0 & 0 & 0 \\
		0 & \mathbf{I}_{\text{E}_1}^{\text{T}} \mathbf{I}_{\text{E}_1} & \mathbf{I}_{\text{E}_1}^{\text{T}} \mathbf{I}_{\text{E}_2} \\ 
		0 & \mathbf{I}_{\text{E}_2}^{\text{T}} \mathbf{I}_{\text{E}_1} & \mathbf{I}_{\text{E}_2}^{\text{T}} \mathbf{I}_{\text{E}_2}
	\end{pmatrix}
\end{equation*}
with their entries defined in \eqref{eq:IJs}.
\end{prop}
\begin{exmp}
Next we will show application of Proposition \ref{Prop2D_1} on solving an eigenvalue problem for a selected geometry. Consider 3-star graph in Figure \ref{fig:2DGraph} with the choice of (clockwise) angles $\delta_1 = \pi/2$ and $\delta_2 = \pi$. In this case the edge's local coordinate basis satisfy
\begin{equation*}
	\vec i_1 = -\vec j_2 = \vec j_3 = \vec E_1 = [1 ~ 0]^{\text{T}}, \qquad
	\vec j_1 = \vec i_2 = -\vec i_3 = \vec E_2 = [0 ~ 1]^{\text{T}}
\end{equation*}
Straightforward calculations shows that $\mathbf{1}^{\text{T}}\mathbf{J}_{\text{E}_1} = 0$, $\mathbf{1}^{\text{T}}\mathbf{J}_{\text{E}_2} = 1$, and moreover
\begin{equation*}
\mathbf{I}_{\text{E}_1}^{\text{T}} \mathbf{I}_{\text{E}_1} = 
\mathbf{J}_{\text{E}_2}^{\text{T}} \mathbf{J}_{\text{E}_2} = 
1, \qquad 
\mathbf{I}_{\text{E}_1}^{\text{T}} \mathbf{I}_{\text{E}_2} = 
\mathbf{J}_{\text{E}_1}^{\text{T}} \mathbf{J}_{\text{E}_2} = 0, \qquad 
\mathbf{I}_{\text{E}_2}^{\text{T}} \mathbf{I}_{\text{E}_2} = 
\mathbf{J}_{\text{E}_1}^{\text{T}} \mathbf{J}_{\text{E}_1} = 2
\end{equation*}
Applying this set of geometric information in \eqref{eq:M2DExp}, the eigenvalue problem is equivalent on finding  ($n$-dependent) coefficient vector $[v_{\vt_c}^\circ, (\vec \omega_\vt^{\hspace{0.5mm}\circ} \cdot \vec E_1),  (\vec \omega_\vt^{\hspace{0.5mm}\circ} \cdot \vec E_2)]^{\text{T}}$ in the kernel of matrix $\mathbb{M}(\lambda)$ of the form
\begin{equation*}
\begin{pmatrix}
3a \phi_3'''(1)+\lambda \mathfrak{m}_{\vt_c} & 0 & -a \phi_4'''(1)\\
0 & 2a\phi_4''(1)+d\psi_2'(1)-\lambda \mathfrak{m}_{\vt_c} & 0 \\
-a \phi_3''(1) & 0 & a\phi_4''(1)+2d\psi_2'(1)-\lambda \mathfrak{m}_{\vt_c}
\end{pmatrix}
\end{equation*}
The fact that $\mathbb{M}(\lambda)$ is singular at an eigenvalue $\lambda \not \in \Sigma^\text{D}$ implies that the spectrum of operator $\cA$ can be further factorized as $\sigma(\cA) = \Sigma^\text{D} \cup \Sigma_1 \cup \Sigma_2$ where $\Sigma^\text{D}$ is defined in \eqref{eq:sigMD} and 
\begin{subequations}
\begin{equation}
	\Sigma_1 := \big\{\lambda \in \bR_+ ~:~ 2a\phi_4''(1)+d\psi_2'(1)-\lambda \mathfrak{m}_{\vt_c} = 0 \big\}
\end{equation}
\begin{equation}
\Sigma_2 := \big\{\lambda \in \bR_+ ~:~ (3a \phi_3'''(1)+\lambda \mathfrak{m}_{\vt_c})(a\phi_4''(1)+2d\psi_2'(1)-\lambda \mathfrak{m}_{\vt_c})-a^2 \phi_3''(1)\phi_4'''(1) = 0 \big\}
\end{equation}
\end{subequations}
Figure \eqref{fig:EP1} plots the first three eigenfunctions $v\up{n}(x)$ for the rigid-vertex model, i.e. $\kappa_{\omega_\eta}^{-1} \rightarrow 0$, and material parameters $d=1, d = 10^5$. In these plots, the colorbar reflects value of torsion $\eta\up{n}(x)$ along each edge. As it is expected, although these plots coincide with the ones reported in \cite{BE21}, but the characteristics equation is different in multiple ways, interested reader may compare Example 5.2. in \cite{BE21} and Proposition \ref{Prop2D_1} above. 
\begin{figure}[ht]
	\centering
	\includegraphics[width=0.975\textwidth]{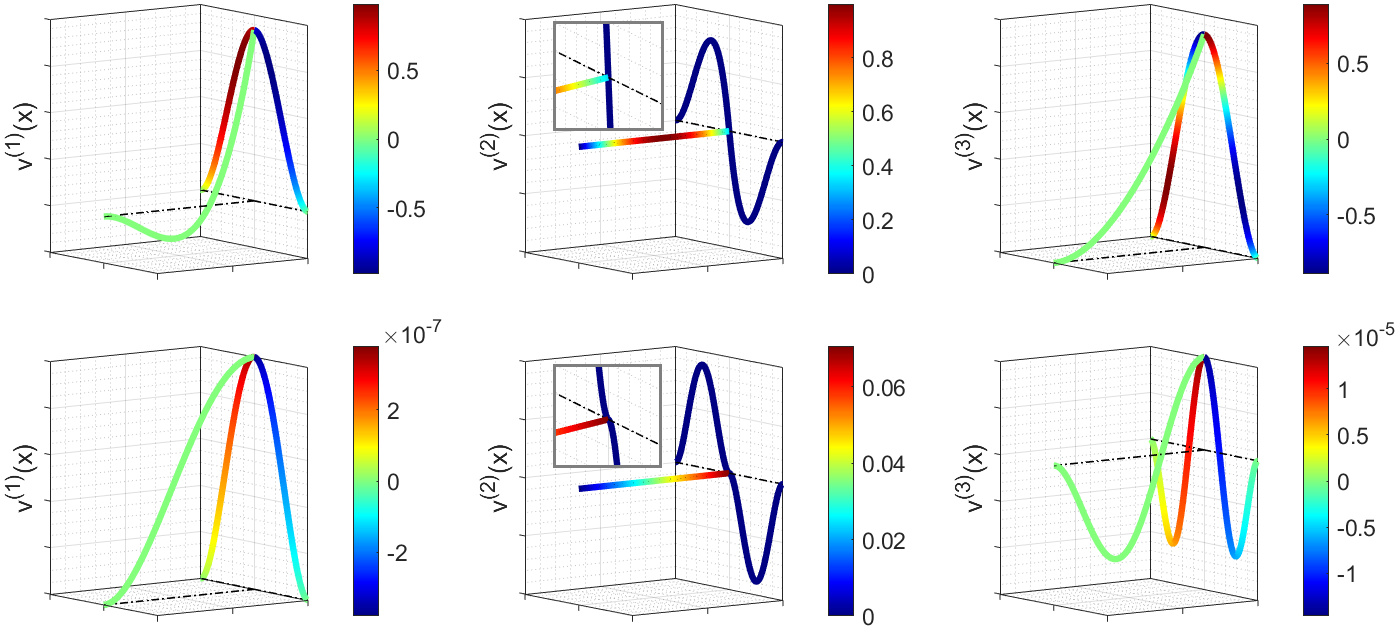}
	\caption{Plots of first three eigenfunction corresponding to the Hamiltonian on 3-star planar graph in global coordinate for rigid-vertex case. The top row is for material parameters $a=d=1$ while the bottom row stands for $a=1, d = 10^5$. The colorbar shows value of angular displacement along the edge.}
	\label{fig:EP1}
\end{figure}
\begin{figure}[ht]
	\centering
	\includegraphics[width=0.975\textwidth]{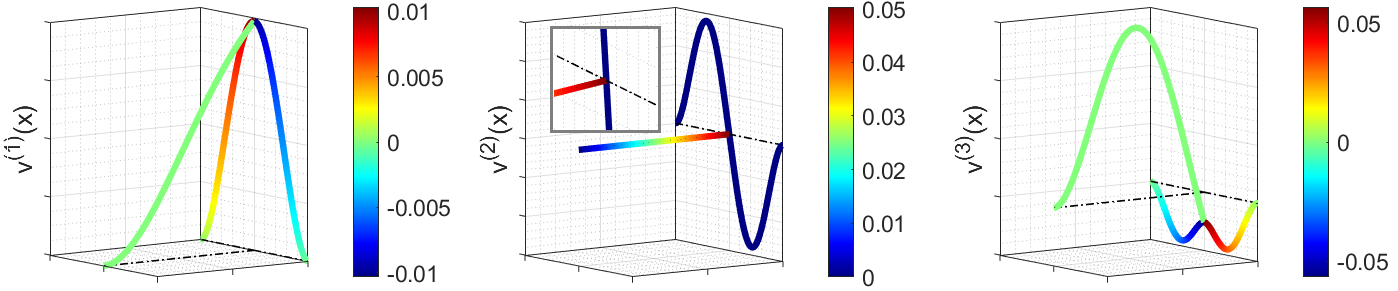}
	\caption{Same as ($a=1, d=10^5$) bottom row of Figure \ref{fig:EP1}, but central vertex is rigid in all it's degrees of freedom except the angular displacement one by setting $\kappa_{\omega_\eta}^{-1} = 1$.}
	\label{fig:EP2}
\end{figure}
\end{exmp}
Figure \ref{fig:EP2} plots the eigenfunctions for materials parameter $a = 1, d=10^5$ while semi-rigid property of central vertex comes to play by setting $\kappa_{\omega_\eta}^{-1} = 1$. While (one-by-one) pictorial comparison of middle pictures in the above Figures may not reveal noticeable different, but in a neighborhood of central vertex $\vt_c$ the behavior is changed considerably, see magnified rectangular windows in plots above on  the deformation's mode in a neighborhood of the central vertex. 

It has been discussed in \cite{BE21} that for the rigid-vertex case, the above results will not match with the formulation \cite{KKU15} in which only scalar field $v(x)$ exists in the Hamiltonian. Taking edge's angular displacement stiffness $d$ to $0$ or to $\infty$ will activate only eigenfunctions $\eta\up{n}(x)$ or suppresses rotation of (with normal vector $\vec E_3$) tangent plane at the central vertex respectively. The former is due to decreasing graph's resistance to the angular displacement while the latter is causes by enforcing small torsion of edges near the vertex due to high resistance of graph to this mode of deformation. Although limit of $d \rightarrow \infty$ is necessary (and not sufficient) on coinciding the two models in \cite{KKU15} and \cite{BE21}, but the key observation here is to (only) relax rigidity of vertex to the angular displacement (torsion) by letting $\kappa_{\omega_\eta} \rightarrow 0$. This makes application of proposed generalized vertex conditions in this manuscript an essential tool for answering problem (ii) stated in the introduction Section. 
\subsection{Joint Model on Fields Decoupling}
Rest of the manuscript will be dedicated to a formal proof on decomposing vector-valued Hamiltonian to a set of scalar ones discussed above. First, observe that for a diagonal form of matrix $\cK_{\omega_e}\up{v}$ in \eqref{primaryVertexCond2H1}, and by an assumption that $\mathfrak{m}_\vt = 0$, then operator $\cA\up{\text{out}}$ reduces to coupled system 
\begin{equation}
\label{eq:outPlane2}
\begin{pmatrix}
v_e\\
\eta_e
\end{pmatrix}
\mapsto
\begin{pmatrix}
a_e v_e''''\\
-d_e \eta_e''
\end{pmatrix}
\end{equation}
acting on functions in the space
$\prod_{e\in \cE} \cH^4(e) \times \prod_{e\in \cE} \cH^2(e)$ satisfying at a vertex $\vt$ with degree $n_\vt$, the primary conditions
\begin{subequations}
	\label{vertexCondPlanarPrimaryH1_2}
	\begin{gather*}
	\label{primaryVertexCond1H1_2}
	v_1 - s_1^\vt a_1 \kappa_{g_v}^{-1} v_1'''  = \cdots = v_{n_\vt}- s_{n_\vt}^\vt a_{n_\vt} \kappa_{g_v}^{-1} v_{n_\vt}''' \\
	\label{primaryVertexCond2H1_2}
	\hspace{2.5mm}(\eta_1+s_1^\vt d_1 \kappa_{\omega_\eta}^{-1} \eta_1') \vec i_1 - (v_1'+s_1^\vt a_1 \kappa_{\omega_v}^{-1} v_1'')\vec j_1 = \cdots = (\eta_3+s_3^\vt d_3 \kappa_{\omega_\eta}^{-1} \eta_3') \vec i_3 - (v_{n_\vt}'+s_{n_\vt}^\vt a_{n_\vt} \kappa_{\omega_v}^{-1} v_{n_\vt}'')\vec j_{n_\vt}
	\end{gather*}
\end{subequations}
along with conjugate ones 
\begin{subequations}
	\label{eq:SS}
	\begin{gather*}
		\label{eq:S1_2}
	s_1^\vt a_1 v_1''' + \cdots + s_{n_\vt}^\vt a_{n_\vt} v_{n_\vt}''' = 0\\
	\label{eq:S2_2}
	\hspace{7.5mm}s_1^\vt(d_1 \eta_1' \vec i_1 - a_1 v_1'' \vec j_1) + \cdots +
	s_{n_\vt}^\vt(d_{n_\vt} \eta_{n_\vt}' \vec i_{n_\vt} - a_{n_\vt} v_{n_\vt}'' \vec j_{n_\vt}) = 0
	\end{gather*}
\end{subequations}
Following Lemma reformulates the second constraint in primary conditions above which will be essential for rest of discussion. 
\begin{lem}
	\label{equivConditionsPlanar}
	At each $\vt$, the primary vertex condition
	\begin{equation}
		(\eta_e+s_e^\vt d_e \kappa_{\omega_\eta}^{-1} \eta_e') \vec i_e - (v_e'+s_e^\vt a_e \kappa_{\omega_v}^{-1} v_e'')\vec j_e = \vec \omega_\vt^{\hspace{0.5mm} \circ}
	\end{equation}
    holds for all $e \sim \vt$ if and only if 
    \begin{subequations}
    \begin{gather}
    \label{rigidPlane_1}
    (\vec j_2 \cdot \vec i_e)\big(v_1'+s_1^\vt a_1 \kappa_{\omega_v}^{-1}v_1''\big) + 
    (\vec j_e \cdot \vec i_1)\big(v_2'+s_2^\vt a_2 \kappa_{\omega_v}^{-1}v_2''\big) + 
    (\vec j_1 \cdot \vec i_2)\big(v_e'+s_e^\vt a_e \kappa_{\omega_v}^{-1}v_e''\big)
    = 0,\\
    \label{rigidPlane_2}
    (\vec j_2 \cdot \vec j_e)\big(v_1'+s_1^\vt a_1 \kappa_{\omega_v}^{-1}v_1''\big) - 
    (\vec j_e \cdot \vec j_1)\big(v_2'+s_2^\vt a_2 \kappa_{\omega_v}^{-1}v_2''\big) + 
    (\vec j_1 \cdot \vec i_2)\big(\eta_e+s_e^\vt d_e \kappa_{\omega_\eta}^{-1}\eta_e'\big)
    = 0.
    \end{gather}
    \end{subequations}
\end{lem}
\begin{proof}[\normalfont \textbf{Proof of Lemma~\ref{equivConditionsPlanar}}] 
	Establishing the above result is based on the proof of Lemma 4.3 in \cite{BE21}. The main difference here is to replace fields with the extended ones introduced in this manuscript, e.g. $v_e'$ in \cite{BE21} will be replaced by $v_e'+s_e^\vt a_e \kappa_{\omega_v}^{-1}v_e''$ and so on. 
\end{proof}
\begin{remark}
	\label{RemarkDeg2}
	Observe that for the case $n_\vt = 2$, condition \eqref{rigidPlane_1} is trivial resulting $v_1'+s_1^\vt a_1 \kappa_{\omega_v}^{-1}v_1'' = v_1'+s_1^\vt a_1 \kappa_{\omega_v}^{-1}v_1''$ and $v_2'+s_2^\vt a_2 \kappa_{\omega_v}^{-1}v_2'' = v_2'+s_2^\vt a_2 \kappa_{\omega_v}^{-1}v_2''$. Thereby without loss of generality we assume that $n_\vt \ge 3$. Recalling definitions of vectors $\vec \omega_e(\vt)$ and $\vec m_e(\vt)$ in \eqref{eq:omega_lin} and \eqref{eq:FeMe} respectively, then condition \eqref{rigidPlane_1} can be expressed as
	\begin{equation}
	\label{eq:C1}
	\big(\big(\bm{\vec i}_1(\vt) + \kappa_{\omega_v}^{-1} \vec m_1(\vt) \times \vec i_1 \big)\times\big(\bm{\vec i}_2(\vt) + \kappa_{\omega_v}^{-1} \vec m_2(\vt) \times \vec i_2 \big)\big) \cdot \big(\bm{\vec i}_e(\vt) + \kappa_{\omega_v}^{-1} \vec m_e(\vt) \times \vec i_e \big) = 0
	\end{equation}
	for all $e \sim \vt$. In fact, setting $w'_e=0$ in the representation of $\vec \omega_e(\vt)$ and $\vec m_e(\vt)$, then \eqref{eq:C1} has a form 
	\begin{equation}
	\label{eq:mid}
	\big(\big(\vec i_1 +(v_1' + s_1^\vt a_1\kappa_{\omega_v}^{-1}v_1'') \vec k\big) \times \big(\vec i_2 +(v_2' + s_2^\vt a_2\kappa_{\omega_v}^{-1}v_2'') \vec k\big)\big) \cdot 
	\big(\vec i_e +(v_e' + s_e^\vt a_e\kappa_{\omega_v}^{-1}v_e'') \vec k\big) = 0
	\end{equation}
	Now, applying properties $(\vec i_1 \times \vec i_2)\cdot \vec i_e = 0$, $(\vec i_1 \times \vec k)\cdot \vec k = 0$, and $(\vec k \times \vec i_2)\cdot \vec k = 0$ in \eqref{eq:mid} implies that 
	\begin{equation*}
	\begin{split}
	(v_1' + s_1^\vt a_1\kappa_{\omega_v}^{-1}v_1'') \underbrace{(\vec k \times \vec i_2)}_{\vec j_2} \cdot \vec i_e +
	(v_2' + s_2^\vt a_2\kappa_{\omega_v}^{-1}v_2'')\underbrace{(\vec i_1 \times \vec k) \cdot \vec i_e}_{-\vec j_1 \cdot \vec i_e ~ = ~\vec j_e \cdot \vec i_1} + (v_e' + s_e^\vt a_e\kappa_{\omega_v}^{-1}v_e'')\underbrace{(\vec i_1 \times \vec i_2) \cdot \vec k}_{\vec i_2 \cdot (\vec k \times \vec i_1) ~=~ \vec i_2 \cdot \vec j_1} = 0
	\end{split}
	\end{equation*}
	which is condition \eqref{rigidPlane_1} as it has been claimed.
\end{remark}
\begin{remark}
	Letting $\vec{\mathbf{m}}_e(\vt) := \vec i_e + \kappa_{\omega_v}^{-1} \vec m_e \times \vec i_e$, observe that \eqref{rigidPlane_1} has another representation of a form 
	\begin{equation}
	\label{eq:localPlanarG}
	\big(\bm{\vec i}_1(\vt) \times \bm{\vec i}_2(\vt) \big) \cdot \bm{\vec i}_e(\vt) + \big(\vec{\mathbf{m}}_1(\vt) \times \vec{\mathbf{m}}_2(\vt) \big) \cdot \vec{\mathbf{m}}_e(\vt) = 0
	\end{equation}
	for all $e \sim \vt$. For the special class of rigid-joints, i.e. the case $\kappa_{\omega_v}^{-1} \rightarrow 0$, \eqref{eq:localPlanarG} reduces to the condition that all vectors $\{\bm{\vec i}_e(\vt)\}_{e \sim \vt}$ lie in the same plane, i.e. $\left( \bm{\vec i}_1(\vt) \times \bm{\vec i}_2(\vt) \right) \cdot \bm{\vec i}_e(\vt) = 0$ for all $e \sim \vt$. This is an important observation that irrespective to the value of vertex's torsional stiffness $\kappa_{\omega_\eta}$, local planar structure of an unreformed graph will be conserved under frame deformation. 
\end{remark}
Now, let introduce
\begin{equation}
\label{eq:defDe}
\mathfrak{D}_e(v,\eta) := \frac{(\vec j_e \cdot \vec j_1)}{(\vec j_1 \cdot \vec i_2)}\big(v_2'+s_2^\vt a_2 \kappa_{\omega_v}^{-1}v_2''\big) -
\frac{(\vec j_2 \cdot \vec j_e)}{(\vec j_1 \cdot \vec i_2)}\big(v_1'+s_1^\vt a_1 \kappa_{\omega_v}^{-1}v_1''\big) - \eta_e
\end{equation}
Observe that by Remark \ref{RemarkDeg2}, the expression \eqref{eq:defDe} is well-defined. This then allows to write condition \eqref{rigidPlane_2} in terms of rotational moment $\eta'_e$ at vertex $\vt$ of a form
\begin{equation}
\label{eq:rotMom}
s_e^\vt d_e \eta_e' = \kappa_{\omega_\eta} \mathfrak{D}_e(v,\eta) 
\end{equation}
Applying \eqref{eq:rotMom} in the second constraint of conjugate condition implies that
\begin{equation}
\label{eq:obsC1}
s_1^\vt a_1 v_1'' \vec j_1 + \cdots + s_{n_\vt}^\vt a_{n_\vt} v_{n_\vt}'' \vec j_{n_\vt} - \kappa_{\omega_\eta}(\mathfrak{D}_1(v,\eta)  \vec i_1 + \cdots + \mathfrak{D}_{n_\vt}(v,\eta)  \vec i_{n_\vt}) = 0
\end{equation}
Since the differential expression for the operator $\cA\up{\text{out}}$ is already in the ``block-diagonal'' form, applying result of Lemma \ref{equivConditionsPlanar} and \eqref{eq:obsC1}, then under an appropriate limit, the operator $\cA\up{\text{out}}$ will be decomposed to two scalar-valued operators. We summarize this result in the following Proposition.
\begin{prop}
	\label{thm:vetaDecm}
	Under a limit $\kappa_{\omega_\eta} \rightarrow 0$, the operator $\cA\up{\text{out}}$ in \eqref{eq:outPlane2} is decomposed as 
	\begin{equation}
	\cA\up{\text{out}} = \cA_v\up{\text{out}} \oplus \cA_\eta\up{\text{out}}
	\end{equation}
	with $\cA_v\up{\text{out}} : v_e
	\mapsto a_e v_e''''$, and satisfying at each (internal) vertex $\vt \in \cV$ with a degree $n_\vt$, conditions
	\begin{subequations}
		\begin{gather}
		\label{dispRigid1_2}
		v_1 - s_1^\vt a_1 \kappa_{g_v}^{-1} v_1'''  = \cdots = v_{n_\vt}- s_{n_\vt}^\vt a_{n_\vt} \kappa_{g_v}^{-1} v_{n_\vt}''', \\
		\label{rotationRigid1_2}
		(\vec j_2 \cdot \vec i_e)(v_1'+s_1^\vt a_1 \kappa_{\omega_v}^{-1}v_1'') + 
		(\vec j_e \cdot \vec i_1)(v_2'+s_2^\vt a_2 \kappa_{\omega_v}^{-1}v_2'') + 
		(\vec j_1 \cdot \vec i_2)(v_e'+s_e^\vt a_e \kappa_{\omega_v}^{-1}v_e'')
		= 0,\\
		\label{eq:S1_3}
		\hspace{4.75mm}s_1^\vt a_1 v_1''' + \cdots + s_{n_\vt}^\vt a_{n_\vt} v_{n_\vt}''' = 0,\\
		\label{eq:S2_3}
		\hspace{7.5mm}s_1^\vt a_1 v_1'' \vec j_1 + \cdots + s_{n_\vt}^\vt a_{n_\vt} v_{n_\vt}'' \vec j_{n_\vt} = 0.
		\end{gather}
	\end{subequations}
	Moreover, $\cA_\eta\up{\text{out}} : \eta_e \mapsto -d_e \eta_e''$ satisfies at each (internal) vertex $\vt \in \cV$ condition  
	\begin{equation}
	d_e \eta_e' = 0.
	\end{equation}
\end{prop}
Similar decomposition result of a form stated in the Proposition \ref{thm:vetaDecm} can be established for $(w,u)$ fields corresponding $\cA\up{\text{in}}$ operator by letting $\kappa_{g_u} \rightarrow 0$. Avoiding detail calculation this result is summarized in the following Proposition.  
\begin{prop}
	\label{thm:wuDecm}
	Under a limit $\kappa_{g_u} \rightarrow 0$, the operator $\cA\up{\text{in}}$ is decomposed as
	\begin{equation}
	\cA\up{\text{in}} = \cA_w\up{\text{in}} \oplus \cA_u\up{\text{in}}
	\end{equation}
	with $\cA_w\up{\text{in}} : w_e
	\mapsto b_e w_e''''$, and satisfying at each (internal) vertex $\vt \in \cV$ with a degree $n_\vt$, conditions
	\begin{subequations}
		\begin{gather}
	     \label{rotationRigid1_3}
	     (\vec j_2 \cdot \vec i_e)(w_1-s_1^\vt b_1 \kappa_{\omega_w}^{-1}w_1''') + 
	     (\vec j_e \cdot \vec i_1)(w_2-s_2^\vt b_2 \kappa_{\omega_w}^{-1}w_2''') + 
	     (\vec j_1 \cdot \vec i_2)(w_e-s_e^\vt b_e \kappa_{\omega_w}^{-1}w_e''')
	     = 0,\\
	     \label{dispRigid1_3}
	     w_1' + s_1^\vt b_1 \kappa_{g_w}^{-1} w_1''  = \cdots = w_{n_\vt} + s_{n_\vt}^\vt b_{n_\vt} \kappa_{g_w}^{-1} w_{n_\vt}'', \\
	     \label{eq:S2_4}
	     \hspace{7.5mm}s_1^\vt b_1 w_1'' \vec j_1 + \cdots + s_{n_\vt}^\vt b_{n_\vt} w_{n_\vt}'' \vec j_{n_\vt} = 0,\\
	     \label{eq:S1_4}
	     \hspace{4.75mm}s_1^\vt a_1 v_1''' + \cdots + s_{n_\vt}^\vt a_{n_\vt} v_{n_\vt}''' = 0.
	    \end{gather}
    \end{subequations}
	Moreover, $\cA_u\up{\text{in}} : u_e \mapsto -c_e u_e''$ satisfies at each (internal) vertex $\vt \in \cV$ condition  
	\begin{equation}
	c_e u_e' = 0.
	\end{equation}
\end{prop}
Combining of Propositions \ref{thm:vetaDecm} and \ref{thm:wuDecm}, thereby we proved the following Theorem, see question (ii) of problem statement in Section \ref{sec:Intro}.
\begin{thm}
	\label{thm:decouplingThm}
	Under limit $(\kappa_{\omega_\eta},\kappa_{g_u}) \rightarrow (0,0)$, vector-valued beam Hamiltonian on planar frames is decomposed as direct sum  
	\begin{equation}
	\cA = \big(\cA_v\up{\text{out}} \oplus \cA_\eta\up{\text{out}}\big) \bigoplus \big(\cA_w\up{\text{in}} \oplus \cA_u\up{\text{in}}\big)
	\end{equation}
	with each scalar-valued Hamiltonian is defined in Propositions \ref{thm:vetaDecm} and \ref{thm:wuDecm}. 
\end{thm}
\begin{remark}
	As a final remark, observe that for a 3-star planar graph and in the limiting case of $(\kappa_{g_v},\kappa_{\omega_v}) \rightarrow (0,0)$, vertex conditions \eqref{dispRigid1_2} are of the form reported for scalar-valued beam operator \cite{KKU15}. Thereby, this special case is a sub-class of the proposed coupled Hamiltonian under certain limits of parameter space of semi-rigid joint derived in this manuscript.  
\end{remark}

\section{Outlook}
\label{sec:Outlook}
Interesting problem is to mathematically investigate the validity of frame model as a structure composed of one-dimensional segments. This could be of interest to dwell on an alternative approach to model junctions between beams which consists in an asymptotic analysis for three dimensional plate models when the thickness tends to zero. There is a significant mathematical literature on this
question for second-order operators see, for example,
\cite{B85,CZ98,Z02,CZ00,BL04,Gri_incol08,Post_book12,MNP13,GM20}, with a variety of operators arising in the limit. Results in this line for the case of fourth-order equations is expected to be of interest to engineering communities working on analysis of 3D structural, e.g. see \cite{D89, LNS19, LNST19}, as well as more theoretically oriented research communities, e.g. see \cite{FJM06, N12, MNP13}. 

Recently, full description of spectra corresponding to the scalar valued fourth-order (Schr{\"o}dinger) operator on a so called class of hexagonal lattices has been discussed in \cite{EH21}. The result of Theorem \ref{thm:decouplingThm} states that scalar-valued Hamiltonian applied in \cite{EH21} is a special case of vector-valued one. Spectral analysis of coupled Hamiltonian on periodic beam lattices equipped with the (general) vertex model proposed in this manuscripts maybe of interest and provides a more complete picture on spectral analysis of such continua, see \cite{FigKuc_siamjam96, MB07,LTWZ19,ES21}.     

\section*{Acknowledgment}
This research was initiated in the Directed Reading Program in the
Department of Mathematics at Texas A\&M University, which is supported
by the NSF grant DMS-1752672.

\section*{Appendix}
\textbf{Derivation of characteristic equation \eqref{eq:MRJ}}. Following Remark \ref{3DRemark_1}, the graph $\Gamma$ is invariant under the following geometric transformations and their products: $\mathbf{R}$ acting as the rotation counterclockwise by $\theta = 2\pi/3$ around the axis $-\vec{i}_0$, and $\mathbf{F}$ acting as the reflection with respect to the plane spanned by $\vec{i}_1$ and $\vec{i}_0$. These transformations generate the group $\cG = D_3$, the dihedral group of degree
$3$, according to the presentation
\begin{equation}
\label{eq:group_presentation}
\cG = \left\langle \mathbf{R}, \mathbf{F} ~:~\mathbf{R}^3 = \mathbf{I}, ~ \mathbf{F}^2 = \mathbf{I}, ~ \mathbf{F} \mathbf{R} \mathbf{F} \mathbf{R} = \mathbf{I} \right\rangle
\end{equation}
Let us now fix the following local bases: take $\vec{j}_1$ to be orthogonal to the plane spanned by $\vec{i}_0$ and $\vec{i}_1$; let $\vec{j}_0=\vec{j}_1$, see Figure~\ref{threeDimGraphExample}; this determines $\vec{k}_0$ and $\vec{k}_1$.  More specifically, 
\begin{subequations}
	\label{eq:4star_bases}
	\begin{gather}
	\vec i_0 = -\vec E_3, \quad \vec j_0 = \vec E_2, \quad \vec k_0 = \vec E_1 \\
	\hspace{7mm}\vec i_1 = \cos(\alpha) \vec E_1 + \sin(\alpha) \vec E_3, \quad \vec j_1 = \vec E_2, \quad \vec k_1 = -\sin(\alpha) \vec E_1 + \cos(\alpha) \vec E_3
	\end{gather}
\end{subequations}
We further assume
\begin{equation}
\label{eq:rotation_fo_basis}
\vec{i}_2 = \mathbf{R}\vec{i}_1,\quad
\vec{j}_2 = \mathbf{R}\vec{j}_1,\quad 
\vec{k}_2 = \mathbf{R}\vec{k}_1,
\quad\mbox{and}\quad
\vec{i}_3 = \mathbf{R}\vec{i}_2,\quad
\vec{j}_3 = \mathbf{R}\vec{j}_2,\quad 
\vec{k}_3 = \mathbf{R}\vec{k}_2.
\end{equation}
With a slight abuse of notation by using the same letters for the matrices realizing, we obtain the following geometric
representation of $\cG$,
\begin{equation}
\label{eq:matrices_R_F}
\mathbf{R} =
\begin{pmatrix}
\cos(\delta)& -\sin(\delta) & \hspace{3mm}0\\
\sin(\delta)& \hspace{3mm}\cos(\delta)& \hspace{3mm}0\\
0 & \hspace{3mm}0 &\hspace{3mm} 1
\end{pmatrix},
\qquad
\mathbf{F} = 
\begin{pmatrix}
1 & \hspace{3mm}0 & \hspace{3mm}0 \\
0 & -1 &\hspace{3mm} 0 \\
0 & \hspace{3mm}0 & \hspace{3mm}1
\end{pmatrix}.
\end{equation}
Next we will discuss application of domain decomposition stated in Theorem \ref{thm:reducing_antenna_tower}. We start with the coordinate decomposition of vector $\vec g_{\vt_c}^{\hspace{0.5mm}\circ}$ in the global coordinate system 
\begin{equation}
\label{eq:gvGlobal_1}
\vec g_{\vt_c}^{\hspace{0.5mm}\circ} = (\vec g_{\vt_c}^{\hspace{0.5mm}\circ} \cdot \vec E_1) \vec E_1 + (\vec g_{\vt_c}^{\hspace{0.5mm}\circ} \cdot \vec E_2) \vec E_2 + (\vec g_{\vt_c}^{\hspace{0.5mm}\circ} \cdot \vec E_3) \vec E_3
\end{equation}
and similarly for vector $\vec \omega_{\vt_c}^{\hspace{0.5mm}\circ}$ as
\begin{equation}
\label{eq:omegavGlobal_1}
\vec \omega_{\vt_c}^{\hspace{0.5mm}\circ} = (\vec \omega_{\vt_c}^{\hspace{0.5mm}\circ} \cdot \vec E_1) \vec E_1 + (\vec \omega_{\vt_c}^{\hspace{0.5mm}\circ} \cdot \vec E_2) \vec E_2 + (\vec \omega_{\vt_c}^{\hspace{0.5mm}\circ} \cdot \vec E_3) \vec E_3
\end{equation}
\textbf{Irreducible Representation} $\mathbf{\cH}_\omega$. 
Stating with the conditions
\begin{equation}
\begin{pmatrix}
\mathbf{R}\vec{g}_0 \\ \mathbf{R}\vec{g}_3 \\ \mathbf{R}\vec{g}_1 \\ \mathbf{R}\vec{g}_2    
\end{pmatrix}
= \omega
\begin{pmatrix}
\vec{g}_0 \\ \vec{g}_1 \\ \vec{g}_2 \\ \vec{g}_3
\end{pmatrix}
\qquad\mbox{and}\qquad
\begin{pmatrix}
\eta_0 \\ \eta_3 \\ \eta_1 \\ \eta_2    
\end{pmatrix}
= \omega
\begin{pmatrix}
\eta_0 \\ \eta_1 \\ \eta_2 \\ \eta_3
\end{pmatrix},
\end{equation}
and by referring to the Theorem \ref{thm:reducing_antenna_tower}, for beam $e_0$ properties
\begin{equation}
\label{eq:w_beam0}
u_0 = 0, \qquad \eta_0 = 0, \qquad
b_0 w_0 = i a_0 v_0,
\end{equation}
holds, while on the other beams
\begin{align}
\label{eq:w_legs}
v_2 &= \omega v_1,
&w_2 &= \omega w_1,
&u_2 &= \omega u_1,
&\eta_2 &= \omega \eta_1,\\
v_3 &= \cc\omega v_1,
&w_3 &= \cc\omega w_1,
&u_3 &= \cc\omega u_1,
&\eta_3 &= \cc\omega \eta_1.
\end{align}
Moreover, in the space $\cH_{\cc{\omega}}$, the two vectors $\vec g_{\vt_c}^{\hspace{0.5mm}\circ}$ and $\vec \omega_{\vt_c}^{\hspace{0.5mm}\circ}$ satisfy $\vec g_{\vt_c}^{\hspace{0.5mm}\circ}\cdot\vec E_3 = 0$ and $ \vec \omega_{\vt_c}^{\hspace{0.5mm}\circ}\cdot\vec E_3 = 0$. Next we will apply the vertex conditions \eqref{eq:semirigidVertexDefn} and \eqref{eq:S} on the appropriate fields. By symmetry reduction in \eqref{eq:w_legs}, we will denote by $u(x) := u_s(x)$ for $s = 1,2,3$, and similarly for the rest of the fields on (leg) edges. Setting $e = e_0$ in \eqref{eq:semiRigid_displacement} and applying the property $u_0 \equiv 0$, then 
\begin{equation}
w_0(\ell_0) \vec j_0 + v_0(\ell_0) \vec k_0 + \kappa_{g_0}^{-1}(-b_0 w_0'''(\ell_0) \vec j_0 -a_0 v_0'''(\ell_0) \vec k_0) = \vec g_{\vt_c}^{\hspace{0.5mm}\circ}
\end{equation}
Representation of local basis in the global coordinate in \eqref{eq:4star_bases} along with expansion \eqref{eq:gvGlobal_1} implies two conditions
\begin{subequations}
	\begin{equation}
	\label{eq:condg0E1_1}
	\hspace{2mm}v_0(\ell_0) - a_0 \kappa_{g_0}^{-1} v_0'''(\ell_0) = \vec g_{\vt_c}^{\hspace{0.5mm}\circ}\cdot\vec E_1, 
	\end{equation}
	\begin{equation}
	w_0(\ell_0) - b_0 \kappa_{g_0}^{-1} w_0'''(\ell_0) = \vec g_{\vt_c}^{\hspace{0.5mm}\circ}\cdot\vec E_2.
	\end{equation}
\end{subequations}
Setting $e = e_1$, then \eqref{eq:semiRigid_displacement} is equivalent to condition
\begin{equation}
u(\ell) \vec i_1 + w(\ell) \vec j_1 + v(\ell) \vec k_1 + \kappa_{g_1}^{-1}(c u'(\ell) \vec i_1 -b w'''(\ell) \vec j_1 - a v'''(\ell) \vec k_1) = \vec g_{\vt_c}^{\hspace{0.5mm}\circ}
\end{equation}
Following similar steps by applying representation of local coordinates in the global one, then one realizes three independent conditions as
\begin{subequations}
	\label{eq:condg0E1_2}
	\begin{gather}
	(u(\ell) + c \kappa_{g_1}^{-1} u'(\ell))\cos(\alpha) - 
	(v(\ell) - a \kappa_{g_1}^{-1} v'''(\ell))\sin(\alpha) = \vec g_{\vt_c}^{\hspace{0.5mm}\circ}\cdot\vec E_1, \\
	w(\ell) -b \kappa_{g_1}^{-1} w'''(\ell) = \vec g_{\vt_c}^{\hspace{0.5mm}\circ}\cdot\vec E_2,\\
	(u(\ell) + c \kappa_{g_1}^{-1} u'(\ell))\sin(\alpha) +
	(v(\ell) - a \kappa_{g_1}^{-1} v'''(\ell))\cos(\alpha) = 0.
	\end{gather}
\end{subequations}
Same types of analysis can be applied to utilize the implication of vertex condition \eqref{eq:semiRigid_rotation}. In fact by setting $e = e_0$, then 
\begin{subequations}
	\begin{equation}
	+w_0'(\ell_0) + b_0 \kappa_{\omega_0}^{-1} w_0''(\ell_0) = \vec \omega_{\vt_c}^{\hspace{0.5mm}\circ}\cdot\vec E_1,
	\end{equation}
	\begin{equation}
	\label{eq:condomega0E1_1}
	\hspace{1mm}-v_0'(\ell_0) - a_0 \kappa_{\omega_0}^{-1} v_0''(\ell_0) = \vec \omega_{\vt_c}^{\hspace{0.5mm}\circ}\cdot\vec E_2.
	\end{equation}
\end{subequations}
Moreover, for edge $e_1$ the vertex condition \eqref{eq:semiRigid_rotation} turns to the following three independent conditions
\begin{subequations}
	\label{eq:condomega0E1_2}
	\begin{gather}
	(\eta(\ell) + d \kappa_{\omega_1}^{-1} \eta'(\ell))\cos(\alpha) - (w'(\ell) + b \kappa_{\omega_1}^{-1} w''(\ell))\sin(\alpha) = \vec \omega_{\vt_c}^{\hspace{0.5mm}\circ}\cdot\vec E_1, \\
	-v(\ell) -a \kappa_{\omega_1}^{-1} v''(\ell)) = \vec \omega_{\vt_c}^{\hspace{0.5mm}\circ}\cdot\vec E_2,\\
	(\eta(\ell) + d \kappa_{\omega_1}^{-1} \eta'(\ell))\sin(\alpha) + (w'(\ell) + b \kappa_{\omega_1}^{-1} w''(\ell))\cos(\alpha) = 0.
	\end{gather}
\end{subequations}
It remains to apply the dynamics of vertex $\vt_c$ in \eqref{eq:S}. Expansion of the net force in \eqref{eq:S1} is equivalent to condition  
\begin{equation}
\label{eq:antena_balance_forces}
\begin{split}
cu'(\ell)(\vec i_1 + \omega \vec i_2 + \bar \omega \vec i_3)
- bw'''(\ell)(\vec j_1 + \omega \vec j_2 + \bar \omega \vec j_3) 
&-av'''(\ell)(\vec k_1 + \omega \vec k_2 + \bar \omega \vec k_3)\\
&-b_0w_0'''(\ell_0) \vec j_0 - a_0v_0'''\vec k_0= \lambda \mathfrak{m}_{\vt_c} \vec g_{\vt_c}^{\hspace{0.5mm}\circ}
\end{split}
\end{equation}
But due to expansion \eqref{eq:4star_bases}, identities
\begin{equation*}
(\cos(\alpha))^{-1}(\vec i_1 + \omega \vec i_2 + \bar \omega \vec i_3) = i(\vec j_1 + \omega \vec j_2 + \bar \omega \vec j_3) = -(\sin(\alpha))^{-1}(\vec k_1 + \omega \vec k_2 + \bar \omega \vec k_3) = \frac{3}{2}(\vec E_1 + i \vec E_2)
\end{equation*}
holds. Expansion of $\vec g_{\vt_c}^{\hspace{0.5mm}\circ}$ in the global coordinate along with factorization of terms in direction of $\vec E_1$ turns to the vertex condition
\begin{equation}
\label{eq:standardB1}
\frac{3}{2}cu'(\ell)\cos(\alpha) + \frac{3}{2}biw'''(\ell)
+ \frac{3}{2}av'''(\ell)\sin(\alpha) - a_0v_0'''(\ell_0) = 
\lambda \mathfrak{m}_{\vt_c} (\vec g_{\vt_c}^{\hspace{0.5mm}\circ}\cdot\vec E_1) 
\end{equation}
We stress that condition corresponding direction $\vec E_2$ is linearly dependent to the one stated in \eqref{eq:standardB1}, see Remark \ref{linearDependt} for details. Application of the vertex condition \eqref{eq:S2} is equivalent to condition 
\begin{align*}
d\eta'(\ell)(\vec i_1 + \omega \vec i_2 + \bar \omega \vec i_3) 
-av''(\ell)(\vec j_1 + \omega \vec j_2 + \bar \omega \vec j_3) 
&+bw''(\ell)(\vec k_1 + \omega \vec k_2 + \bar \omega \vec k_3) \\
&-a_0v_0''(\ell_0)\vec j_0 + b_0w_0''(\ell_0) \vec k_0 = \lambda \mathfrak{m}_{\vt_c} \vec \omega_{\vt_c}^{\hspace{0.5mm}\circ}
\end{align*}
which by following similar steps in derivation of \eqref{eq:standardB1} is reduces to a single independent condition 
\begin{equation}
\label{standardB2}
\frac{3}{2}d i\eta'(\ell) \cos(\alpha)
- \frac{3}{2}biw''(\ell)\sin(\alpha)
- \frac{3}{2}av''(\ell)- a_0v_0''(\ell_0) = \lambda \mathfrak{m}_{\vt_c} (\vec \omega_{\vt_c}^{\hspace{0.5mm}\circ}\cdot\vec E_2)
\end{equation}
With the restrictions given by \eqref{eq:w_beam0} and
\eqref{eq:w_legs}, the eigenvalue problem is
fully determined by functions $v, w, u$ and $\eta$ on the base edges and the function $v_0$ defined on vertical edge. Following Remark will identify set of linearly independent conditions out of the ones have been derived above.  
\begin{remark}
	\label{linearDependt}
	Applying the relations $\vec g_{\vt_c}^{\hspace{0.5mm}\circ}\cdot\vec E_2 = +\mathrm{i}(\vec g_{\vt_c}^{\hspace{0.5mm}\circ}\cdot\vec E_1)$ and $\vec \omega_{\vt_c}^{\hspace{0.5mm}\circ}\cdot\vec E_1 = - \mathrm{i}(\vec \omega_{\vt_c}^{\hspace{0.5mm}\circ}\cdot\vec E_2)$ for subspace $\cH_\omega$ stated in the Theorem \ref{thm:reducing_antenna_tower}, then set of linearly independent conditions is equivalent to \eqref{eq:condg0E1_1}, \eqref{eq:condg0E1_2}, \eqref{eq:condomega0E1_1}, \eqref{eq:condomega0E1_2}, along with net force and moment conditions in \eqref{eq:standardB1} and \eqref{standardB2} respectively.
\end{remark}

\section*{References}

\bibliographystyle{abbrv}
\bibliography{ref}

\end{document}